\documentclass[12pt]{iopart}
\usepackage{latexsym}
\usepackage{amscd}
\begin{document}

 \def\lambdabar{\protect\@lambdabar}
\def\@lambdabar{%
\relax
\bgroup
\def\@tempa{\hbox{\raise.73\ht0
\hbox to0pt{\kern.25\wd0\vrule width.5\wd0
height.1pt depth.1pt\hss}\box0}}%
\mathchoice{\setbox0\hbox{$\displaystyle\lambda$}\@tempa}%
{\setbox0\hbox{$\textstyle\lambda$}\@tempa}%
{\setbox0\hbox{$\scriptstyle\lambda$}\@tempa}%
{\setbox0\hbox{$\scriptscriptstyle\lambda$}\@tempa}%
\egroup
}

\def\bbox#1{%
\relax\ifmmode
\mathchoice
{{\hbox{\boldmath$\displaystyle#1$}}}%
{{\hbox{\boldmath$\textstyle#1$}}}%
{{\hbox{\boldmath$\scriptstyle#1$}}}%
{{\hbox{\boldmath$\scriptscriptstyle#1$}}}%
\else
\mbox{#1}%
\fi
}
\def\msf{\hbox{{\sf M}}}
\def\psf{\hbox{{\sf P}}}
\def\Nsf{\hbox{{\sf N}}}
\def\Tsf{\hbox{{\sf T}}}
\def\Asf{\hbox{{\sf A}}}
\def\Bsf{\hbox{{\sf B}}}

\def\msfsim{\bbox{{\sf M}}_{\scriptstyle\rm(sym)}}
\newcommand{\mcsim}{ {\sf M}_{ {\scriptstyle \rm {(sym)} } i_1\dots i_n}}
\newcommand{\mcs}{ {\sf M}_{ {\scriptstyle \rm {(sym)} } i_1i_2i_3}}

\newcommand{\beqan}{\begin{eqnarray*}}
\newcommand{\eeqan}{\end{eqnarray*}}
\newcommand{\beqa}{\begin{eqnarray}}
\newcommand{\eeqa}{\end{eqnarray}}

 \newcommand{\suml}{\sum\limits}
\newcommand{\intl}{\int\limits}
\newcommand{\rvec}{\bbox{r}}
\newcommand{\xivec}{\bbox{\xi}}
\newcommand{\Avec}{\bbox{A}}
\newcommand{\Rvec}{\bbox{R}}
\newcommand{\Evec}{\bbox{E}}
\newcommand{\Bvec}{\bbox{B}}
\newcommand{\Svec}{\bbox{S}}
\newcommand{\avec}{\bbox{a}}
\newcommand{\nablav}{\bbox{\nabla}}
\newcommand{\nuvec}{\bbox{\nu}}
\newcommand{\bvec}{\bbox{\beta}}
\newcommand{\vvec}{\bbox{v}}
\newcommand{\jvec}{\bbox{j}}
\newcommand{\nvec}{\bbox{n}}
\newcommand{\pvec}{\bbox{p}}
\newcommand{\mvec}{\bbox{m}}
\newcommand{\evec}{\bi{e}}
\newcommand{\eps}{\varepsilon}
\newcommand{\la}{\lambda}
\newcommand{\rad}{\mbox{\footnotesize rad}}
\newcommand{\scr}{\scriptstyle}
\newcommand{\latens}{\bbox{\Lambda}}
\newcommand{\pitens}{\bbox{\Pi}}
\newcommand{\cm}{{\cal M}}
\newcommand{\cp}{{\cal P}}

\renewcommand{\d}{\partial}
\def\rmi{{\rm i}}
\def\rme{\hbox{\rm e}}
\def\rmd{\hbox{\rm d}}

\title{ Expressing the power radiated by electric charged systems}
\author{C. Vrejoiu and D. Nicmoru\c{s}}
\address{Faculty of Physics, University of Bucharest, 76900, 
 Bucharest-Magurele, Romania E-mail : cvrejoiu@yahoo.com, diananicmorus@
 hotmail.com}
\begin{abstract}
After a systematic introduction of some formulae for the energy radiated by  
localized electric charges and currents distributions, one considers  the multipole 
radiation and the reduction of the multipole tensors to the symmetric traceless 
ones.
\end{abstract}
\section{Introduction}
In the calculation of the energy radiated at large distances by a localised electric
charged system it is not necessary to know the exact expressions of the electromagnetic
 fields $\Evec$ and $ \Bvec$ or of the potentials $\Avec$ and $\Phi$. One may avoid
 the exact calculation, sometimes relatively complicate, in a simple way based on a
 formula for the power radiated by a charged system described by the charge $\rho$ and current
 $\jvec$ densities with supports included in a finite domain $\cal D$ \cite{land}:
\begin{equation}\label{1}
\frac{\rmd P}{\rmd \Omega}(\nuvec,t)=\frac{r^2}{\mu_0c}\left[ \nuvec\times
\frac{\partial}{\partial t}\Avec_{\rad}(\rvec,t)\right]^2.
\end{equation}
Here the origin $O$ of the coordinates is chosen in the domain ${\cal D}$,
$\nuvec=\rvec/r$,  $\rmd P/\rmd \Omega$ is related to the
flow of the energy detected in the observation point $\rvec$ at large distance $r$ compared with the
dimensions of the given charged system.
The vector $\Avec_{\rad}$ is obtained from the retarded potential
$$\Avec(\rvec,t)=\frac{\mu_0}{4\pi}\intl_{\cal D}\frac{1}{R}\jvec(\rvec',t-\frac{R}{c})\rmd^3x',$$
with $\Rvec=\rvec-\rvec'$, by retaining only the dominant terms at large distances.
A first approximation for this vector is obtained by retaining only the dominant term
$1/r$ from the series expansion of $1/R$,
 $$\frac{1}{R}=\frac{1}{r}+\rvec'\cdot\left(\nablav'\frac{1}{R}\right)_{\rvec'=0}+\dots=\frac{1}{r}
-\rvec'\cdot\nablav\frac{1}{r}+\dots=\frac{1}{r}+\frac{\rvec'\cdot\rvec}{r^3}+\dots=
\frac{1}{r}+O(1/r^2),$$
with a corresponding definition
\begin{equation}\label{2}
\widetilde{\Avec}_{\rad}(\rvec,t)=\frac{\mu_0}{4\pi}\frac{1}{r}\intl_{\cal D}\jvec(\rvec',t-\frac{R}{c})\rmd^3x'.
\end{equation}
In these relations and in the following ones we denote by $O(x^n)$ a series of powers of $x$ beginning with $x^n$.
\par Supposing $r>>\lambda$, where $\lambda$ is an arbitrary wave length from the radiation spectrum, such that  the
observation point is in the wave region, and retaining in equation \eref{1} only the terms having
nonzero limits for $r\to \infty$, one obtains an approximate expression related to the energy flow
observed in the point $\rvec$ and moment $t$. Rigorously, this is that part of the energy flowing in the
neighbourhood of the observation point which contributes to the radiated energy. In the following we
assume to work in this wave region.
\section{The radiation field}
\par In \cite{land}  the equation \eref{1} is justified using the supposed plane wave behaviour of the
radiated field but also in \cite{land}, in a footnote of the page 229, a rigorous proof is suggested
for this. Indeed, this may be done by  considering consistently only the terms from $\Evec$ and $\Bvec$
contributing to the radiation
\cite{vre}.
\par Denoting by $t'=t-R/c$ the retarded time, and by
$$[\rho]=\rho(\rvec',t'),\;\;[\jvec]=\jvec(\rvec',t')$$
the retarded charge and current densities, we have
\beqan
 \Bvec(\rvec,t)&=&\nablav\times\Avec(\rvec,t)=\frac{\mu_0}{4\pi}\nablav\times\left[\frac{1}{r}\intl_{\cal D}
[\jvec]\rmd^3x'\right]+O(1/r^2)\\
 &=&\frac{\mu_0}{4\pi}\left[(\nablav\frac{1}{r})\times\intl_{\cal D}[\jvec]\rmd^3x'-\frac{1}{r}\intl_{\cal D}
\frac{\d }{\d t}[\jvec]\times\nablav t'\rmd^3x'\right]+O(1/r^2)\\
&=&\frac{\mu_0}{4\pi}\frac{1}{r}\intl_{\cal D}(\nablav t')
\times\frac{\d }{\d t}[\jvec]\rmd^3x'+O(1/r^2).
\eeqan
Considering the series expansion of $R$,
$$R=r+\rvec'\cdot(\nablav'R)_{\rvec'=0}+\dots=r-\rvec'\cdot\nablav r+\dots=r-\frac{\rvec\cdot\rvec'}{r}+
O(1/r)$$
we have
$$t'=t-\frac{R}{c}=t-\frac{r}{c}+\frac{1}{c}\nuvec\cdot\rvec'+ O(1/r),$$
and we may write
\begin{equation}\label{3}
\nablav t'=-\frac{1}{c}\nuvec+\frac{\rvec\cdot\rvec'}{c}\nablav\frac{1}{r}+\frac{1}{c}\frac{\rvec'}{r}+\dots
=-\frac{1}{c}\nuvec+O(1/r).
\end{equation}
\par The vector $\Bvec(\rvec,t)$ may be written as
\begin{equation}\label{4}
\fl \Bvec(\rvec,t)=\frac{\mu_0}{4\pi}\frac{1}{cr}\intl_{\cal D}\frac{\d}{\d t}[\jvec]\times\nuvec\rmd^3x'+
O(1/r^2)=\frac{1}{c}\left(\frac{\d\widetilde{\Avec}_{\rad}}{\d t}\times\nuvec\right)+O(1/r^2)
\end{equation}
so that the part of $\Bvec$ contributing to the radiation is, in a first evaluation,
\begin{equation}\label{5}
\widetilde{\Bvec}_{\rad}=\frac{1}{c}\left(\frac{\d\widetilde{\Avec}_{\rad}}{\d t}\times \nuvec\right).
\end{equation}
The electric field $\Evec=-\nablav\Phi-\d\Avec/\d t$ with the retarded scalar potential
$$\Phi(\rvec,t)=\frac{1}{4\pi\eps_0}\intl_{\cal D}\frac{\rho(\rvec',t-R/c)}{R}\rmd^3x'$$
is given by
\beqan
\Evec(\rvec,t)&=&-\frac{1}{4\pi\eps_0}\nablav\left[\frac{1}{r}\intl_{\cal D}[\rho]\rmd^3x'\right]-
\frac{\mu_0}{4\pi}\frac{1}{r}\intl_{\cal D}\frac{\d}{\d t}[\jvec]\rmd^3x'+O(1/r^2)\\
&=& \frac{1}{4\pi\eps_0c}\frac{1}{r}\intl_{\cal D}\frac{\d[\rho]}{\d t}\nuvec\rmd^3x'-\frac{\mu_0}{4\pi}
\frac{1}{r}\intl_{\cal D}\frac{\d[\jvec]}{\d t}\rmd^3x'+O(1/r^2)
\eeqan
where the equation \eref{3} is considered. Writing the continuity equation in the point $\rvec'$ at the retarded
time $t-R/c$,
$$\frac{\d}{\d t}\rho(\rvec',t-\frac{R}{c})+\left[\nablav'\cdot\jvec(\rvec',\tau)\right]_{\tau=t-R/c}=0,$$
and the relations
\beqan
\nablav'\jvec(\rvec',t-\frac{R}{c})&=&\left[\nablav'\jvec(\rvec',\tau)\right]_{\tau=t-R/c}+\frac{\d}{\d t}
\jvec(\rvec',t-\frac{R}{c})\cdot\nablav'(t-\frac{R}{c})\\
&=&\left[\nablav'\jvec(\rvec',\tau)\right]_{\tau=t-R/c}+\frac{1}{c}\nuvec\cdot\frac{\d}{\d t}\jvec(\rvec',t-
\frac{R}{c})+O(1/r),
\eeqan
we have
\beqan
\Evec(\rvec,t)&=&\frac{1}{4\pi\eps_0c}\frac{1}{r}\intl_{\cal D}\nuvec\cdot\nablav'[\jvec]\rmd^3x'+\frac{1}
{4\pi\eps_0c^2}\frac{1}{r}\intl_{\cal D}\left(\nuvec\cdot\frac{\d[\jvec]}{\d t}\right)\nuvec\rmd^3x'\\
&-&\frac{\mu_0}{4\pi}\frac{1}{r}\intl_{\cal D}\frac{\d[\jvec]}{\d t}\rmd^3x'+O(1/r^2).
\eeqan
The first integral in the right hand side of the last equation is zero because $\jvec=0$ on the surface of
${\cal D}$. Because $(\nuvec\cdot\d[\jvec]/\d t)\nuvec-\d[\jvec]/\d t=\nuvec\times(\nuvec\times\d[\jvec]/\d t)$,
we obtain
$$\Evec(\rvec,t)=\frac{\mu_0}{4\pi}\frac{1}{r}\intl_{\cal D}\nuvec\times\left(\nuvec\times\frac{\d[\jvec]}{\d t}
\right)\rmd^3x'+O(1/r^2).$$
As in the case of  the magnetic field, we may write, in a first evaluation, the part
of $\Evec$ contributing to the radiation:
\begin{equation}\label{6}
\widetilde{\Evec}_{\rad}=\nuvec\times\left(\nuvec\times\frac{\d}{\d t}\widetilde{\Avec}_{\rad}\right).
\end{equation}
 The equations verified by the fields $\widetilde{\Evec}_{\rad}$ and $\widetilde{\Bvec}_{\rad}$,
$$\widetilde{\Evec}_{\rad}=c\widetilde{\Bvec}_{\rad}\times\nuvec,\;\;\widetilde{\Bvec}_{\rad}=\frac{1}{c}\nuvec\times
\widetilde{\Evec}_{\rad},\;\;\eps_0\widetilde{\Evec}^2_{\rad}=\frac{1}{\mu_0}\widetilde{\Bvec}^2_{\rad},$$
indicate the plane wave-like local structure of the radiation field corresponding to the radial direction $\nuvec$
of propagation (these relations are valid also for the radiated fields $\Evec_{rad}$ and
$\Bvec_{rad}$ obtained from $\Evec$ and $\Bvec$ by retaining only the terms with $1/r$).
\par Supposing that all the field variables are real functions, the Poynting vector of the radiated field is
$$\Svec_{\rad}=\frac{1}{\mu_0}\Evec_{\rad}\times\Bvec_{\rad}=\eps_0\Evec^2_{\rad}=
\frac{1}{\mu_0}\Bvec^2_{\rad}$$
or
$$\Svec_{\rad}=\frac{1}{\mu_0c}\left(\nuvec\times\frac{\d}{\d t}\Avec_{\rad}\right)^2\,\nuvec.$$
Let the sphere of the radius $r$ with the center in $O$. The radiated energy $\delta\Delta W_{\rad}$
passing through the surface element $\Delta\sigma$, centered on the point $\rvec$ and corresponding to
the solid angle $\Delta\Omega$,  in the time interval $(t,\,t+\delta t)$ is defined by
\begin{equation}\label{7}
\delta\Delta W_{\rad}=\nuvec\cdot\Svec_{\rad}\,r^2\Delta\Omega\delta t.\end{equation}
From the last equation one sees that the angular distribution of the radiation power is given by the
equation \eref{1}.

\section{The radiation of the point electric charge}
Usually, one derives the angular distribution of the power radiated by a point
electric charge $q$ using the results for the fields $\Evec$ and $\Bvec$ obtained
from the Li\'{e}nard-Wiechert potentials and retaining from  the corresponding
expressions only the terms contributing to the radiation \cite{land}-\cite{jack}. Here we illustrate the
simplicity of the calculation using in this case the formula \eref{1}.
 \par For the sake of completeness  we remind  a  concise introduction of the Li\'{e}nard-Wiechert potentials \cite
{beck2}. Considering that the motion of the point charge $q$ is given by the law $t\longrightarrow \xivec(t)$,
 the charge and current densities are represented as
$$\rho(\rvec,t)=q\delta[\rvec-\xivec(t)],\;\;\jvec(\rvec,t)=q\vvec(t)\delta[\rvec-\xivec(t)]$$
where $\delta$ is the Dirac function. The retarded vector potential $\Avec$ is  represented by the
integral
\beqan
\fl \Avec(\rvec,t)&=&\frac{\mu_0}{4\pi}\int\frac{\jvec(\rvec',t-|\rvec-\rvec'|/c)}{|\rvec-\rvec'|}\rmd^3x'
 =\frac{\mu_0}{4\pi}\int\rmd^3x'\intl^{+\infty}_{-\infty}\rmd t'\frac{\jvec(\rvec',t')}{|\rvec-\rvec'|}
\delta[t'-t+|\rvec-\rvec'|/c\, ]\\
\fl &=&\frac{\mu_0q}{4\pi}\intl^{+\infty}_{-\infty}\rmd t'\frac{\vvec(t')}{R(t')}
\delta\left[t'-t+\frac{R(t')}{c}\right]
\eeqan
where $\Rvec(t)=\rvec-\xivec(t)$ gives the position of the observation point with respect to the particle at
the moment $t$.
In the last equation we may use the relation
$$\delta[f(x)]=\suml^n_{i=1}\frac{\delta(x-x_i)}{|f'(x_i)|}$$
supposing  that the equation $f(x)=0$ has $n$ roots $x_1,\dots, x_n$ with $f'(x_i)=(\rmd f/\rmd x)_{x=x_i}$.
\par Here
\begin{equation}\label{8} f(t')=t'-t+\frac{R(t')}{c}\end{equation}
and
\begin{equation}\label{9}f'(t')=1-\bvec(t')\cdot\nvec(t')\end{equation}
where $\bvec=\vvec/c$ and $\nvec(t)=\Rvec(t)/R(t)$. Because $v\;<\;c$ we have $f'(t')\;>\;0$ so that
the equation $f(t')=0$ has only one root $\tau$:
\begin{equation}\label{10}
t-\tau-\frac{1}{c}R(\tau)=0.\end{equation}
Denoting $\eta(t)=1-\bvec(t)\cdot\nvec(t)$, $s(t)=\eta(t)R(t)$, the vector potential $\Avec$ is given by the
well-known Li\'{e}nard-Wiechert expression
\begin{equation}\label{11}
\Avec(\rvec,t)=\frac{\mu_0q}{4\pi}\left(\frac{\vvec}{s}\right)_{\tau},\;\; t-\tau+\frac{R(\tau)}{c}=0,
\end{equation}
and, obviously now,
$$\Phi(\rvec,t)=\frac{q}{4\pi\eps_0}\frac{1}{s(\tau)}.$$
 Now let the potential $\widetilde{\Avec}_{\rad}$ corresponding to the field of the particle
$$\widetilde{\Avec}_{\rad} (\rvec,t)=\frac{\mu_0}{4\pi}\frac{1}{r}\int\jvec(\rvec',t-R/c)\rmd^3x'=\frac{\mu_0}
{4\pi}\frac{1}{r}\intl^{+\infty}_{-\infty}\rmd t'\vvec(t')\delta\left[t'-t+\frac{R(t')}{c}\right]$$
such that
\begin{equation}\label{12}
\widetilde{\Avec}_{\rad} (\rvec,t)=\frac{\mu_0 q}{4\pi}\frac{\vvec(\tau)}{\eta(\tau)r}
 \end{equation}
with $\tau$ defined by the equation \eref{10}. The time derivative of
$\widetilde{\Avec}_{\rad}$ is obtained by a
 simple calculation:
$$\frac{\d\widetilde{\Avec}_{\rad}}{\d t}=\frac{\mu_0q}{4\pi r}\left[\frac{\avec}{\eta}-\frac{\vvec}{\eta^2}
\frac{\d\eta}{\d t}\right]_{\tau}\frac{\d \tau}{\d t}$$
where  $\avec$ is the particle acceleration. To obtain $\d \tau/\d t$ we consider the derivative
of the equation \eref{10}
$$\frac{\d \tau}{\d t}=1+\frac{\d \tau}{\d t}\left(\frac{\vvec\cdot\Rvec}{cR}\right)_{\tau}$$
so that
 we get
\begin{equation}\label{13}
\frac{\d \tau}{\d t}=\left(\frac{1}{1-\nvec\cdot\bvec}\right)_{\tau}=\frac{1}{\eta_0},\end{equation}
 where $\eta_0=\eta(\tau)$.
So
$$
\frac{\d\widetilde{\Avec}_{\rad}}{\d t}=\frac{\mu_0 q}{4\pi r}\left[\frac{\avec}{\eta}-\frac{\vvec}{\eta^2}\left(
-\frac{\avec\cdot\Rvec}{cR}+\frac{v^2}{cR}+\frac{\vvec\cdot\nvec}{R^2}\right)\right]_{\tau}\frac{1}{\eta_0}$$
and we may write
\begin{equation}\label{14}
\frac{\d\widetilde{\Avec}_{\rad}}{\d t}=\frac{\mu_0q}{4\pi}\frac{1}{r\eta^3_0}\left[\eta\avec+(\avec\cdot\nvec)
\bvec\right]_{\tau}+O(1/r^2).\end{equation}
We have to calculate  the expression $\nuvec\times\d\widetilde{\Avec}_{\rad}/\d t$ in order to introduce it in the equation
\eref{1}. The observation point being given by its position vector  $\rvec$, it is clear that in the approximation
considered here the same result is obtained for any chosen origin $O$ inside the domain ${\cal D}$.
The unit vector $\nuvec$
may be replaced by the unit vector $\nvec$ of the direction particle-observation point without changing
the result for the angular distribution of the radiated power. This happens because
\beqan
\nvec=\frac{\Rvec}{R}=\nuvec-(\xivec\cdot\nablav)\frac{\rvec}{r}+\dots
=\nuvec-\frac{\xivec}{r}+\frac{(\xivec\cdot\nuvec)\nuvec}{r}+\dots=\nuvec+O(1/r)\eeqan
and
$$\nuvec\times\frac{\widetilde{\Avec}_{\rad}}{\d t}=\nvec\times\frac{
\widetilde{\d\Avec}_{\rad}}{\d t}+O(1/r^2).$$

By a straightforward calculation one obtains
\begin{equation}\label{15}
\fl\left(\nvec\times\frac{\d\widetilde{\Avec}_{\rad}}{\d t}\right)^2=\left(\frac{\mu_0q}{4\pi}\right)^2
\frac{1}{r^2\eta^6_0}\left[\eta a^2+2\eta(\nvec\cdot\avec)(\bvec\cdot\avec)-(1-\beta^2)(\nvec\cdot\avec)^2
\right]_{\tau}+O(1/r^3)\end{equation}
and with the equation \eref{1} one gets the equation (73,9) given in \cite{land}:
\begin{equation}\label{16}
\rmd P=\frac{q^2}{16\pi^2\eps_0c^3}\frac{1}{\eta^6}
\left[\eta a^2+2\eta(\nvec\cdot\avec)(\bvec\cdot\avec)-(1-\beta^2)(\nvec\cdot\avec)^2
\right]\rmd\Omega\end{equation}
where all the particle parameters are considered at the retarded moment.
Inserting the equation \eref{15} in the equation \eref{7}, and casting the terms of O(1/r), we obtain that part
of the energy passing through the surface element $\Delta \sigma$, in the time interval $\delta t$, which
contributes to the radiation. This is just the radiated part of the energy flowing in the solid angle $\Delta\Omega$
between the moments $\tau$ and $t$. But this energy was emitted by the particle in the
time interval $\delta \tau$ corresponding to the observation interval $\delta t$. Therefore, writing
$$\delta\Delta W_{\rad}=\Svec_{\rad}\cdot\nvec r^2\Delta \Omega\frac{\d t}{\d \tau}\delta \tau,$$
the factor multiplying $\delta \tau$ in the  right hand side of the last equation represents in fact the
radiated part
of the energy emitted by the particle in the unit of time (at an arbitrary time) in the direction $\nvec$
and in the solid angle $\Delta \Omega$. Because $\d t/\d\tau=\eta(\tau)$ we may write the final result for
 the angular distribution of the radiation power emitted by the particle as an expression
differing from the right hand side of the equation \eref{16} by the factor $1/\eta^5$ instead of
 $1/\eta^6$ \cite {land}.

 \section{Multipolar expansion of $\Avec_{rad}$}
The series expansion of the integrand from the equation \eref{2},
by retaining only the $1/r$ terms contributing to the radiation, leads
finally to the multipole expansion of the radiation field.
\par Let us the Taylor series expansion of a function $f(R)$,
\begin{equation}\label{17}
f(R)=\suml^\infty_{n=0}\frac{(-1)^n}{n!}x'_{i_1}\dots x'_{i_n}\d_{
i_1\dots i_n}f(r)=\suml^\infty_{n=0}\frac{(-1)^n}{n!}\rvec'^n||
\nablav^nf(r)
\end{equation}
where
$$\d_{i_1\dots i_n}=\frac{\d}{\d x_{i_1}}\dots \frac{\d}{\d x_{i_n}}$$
and $\avec^n$ is the n-fold tensorial product $(\avec\otimes\dots\otimes
\avec)_{i_1\dots i_n}=a_{i_1}\dots a_{i_n}$. Denoting by
 $\bbox{\sf{T}}^{(n)}$ an $n$th order tensor,
$\bbox{\sf{A}}^{(n)}||\bbox{\sf{B}}^{(m)}$ is an $|n-m|$th 
order tensor with the components
\begin{displaymath}\left(\bbox{\sf{A}}^{(n)}||\bbox{\sf{B}}^{(m)}\right)_{i_1\dots i_{|n-m|}}=
\left\{\begin{array}{ll}A_{i_1\dots i_{n-m}j_1\dots j_m}B_{j_1\dots j_m}& \textrm{,  $n>m$}\\
A_{j_1\dots j_n}B_{j_1\dots j_n}&\textrm{,  $n=m$}\\
A_{j_1\dots j_n}B_{j_1\dots j_ni_1\dots i_{m-n}}&\textrm{,  
$n<m$}\end{array}\right. .\end{displaymath}
By introducing the expansion \eref{17} into the equation \eref{2}, we obtain
\beqa\label{18}
\widetilde{\Avec}_{rad}(\rvec,t)&=&\frac{\mu_0}{4\pi r}\evec_i\suml^\infty_{n=0}\frac{(-1)^n}{n!}\d_{i_1\dots i_n}\intl_
{\cal D}x'_{i_1}\dots x'_{i_n}\, j_i(\rvec',t-\frac{r}{c})\rmd^3x'\nonumber\\
&=&\frac{\mu_0}{4\pi r}
\evec_i\suml^\infty_{n=0}\frac{(-1)^n}{n!}a^{(n)}_i\eeqa
where
\begin{equation}\label{19}
a^{(n)}_i=\d_{i_1\dots i_n}\intl_{\cal D}x'_{i_1}\dots x'_{i_n}\,j_i(\rvec',t-\frac{r}{c})\rmd^3x'\end{equation}
and $\evec_i$ are the orthogonal unit vectors along the axes.
\par In the following we use a generalisation to the dynamic case of a procedure
given in \cite{castel} in the magnetostatic case. Let the identity
$$\nablav[x_i\,\jvec(\rvec,t)]=j_i(\rvec,t)+x_i\nablav\jvec(\rvec,t).$$
Considering the continuity equation $\nablav\jvec+\d\rho/\d t=0$ we may write
\begin{equation}\label{20}
j_i(\rvec,t)=\nablav[x_i\jvec(\rvec,t)]+x_i\frac{\d}{\d t}\rho(\rvec,t)
\end{equation}
and using this last equation in equation \eref{19},we get
\beqan
\fl a_i^{(n)}&=&\d_{i_1\dots i_n}\intl_{\cal D}x'_{i_1}\dots x'_{i_n}\nablav'[x'_i\,
\jvec(\rvec',t_0)]\rmd^3x'+\d_{i_1\dots i_n}\intl_{\cal D}x'_{i_1}\dots x'_{i_n}
x'_i\frac{\d}{\d t}\rho(\rvec',t_0)\rmd^3x'\\
\fl&=&-\d_{i_1\dots i_n}\intl_{\cal D}x'_i\,\jvec(\rvec',t_0)\cdot\nablav'(x'_{i_1}
\dots x'_{i_n})\rmd^3x'+
\d_{i_1\dots i_n}\intl_{\cal D}x'_{i_1}\dots x'_{i_n}x'_i\frac{\d}{\d t}
\rho(\rvec',t_0)\rmd^3x'
\eeqan
denoting $t_0=t-r/c$ and considering a nul surface term because $\jvec=0$ on
$\d{\cal D}$.
 Because of the symmetry of the derivative tensor and introducing the $n$th order
 electric multipole tensor
 \begin{equation}\label{21}
\psf^{(n)}(t)=\intl_{\cal D}\rvec^n\,\rho(\rvec,t)\rmd^3x,\end{equation}
we may write
\beqan
\fl a^{(n)}_i&=&-n\d_{i_1\dots i_n}\int_{\cal D}x'_{i_1}\dots x'_{i_{n-1}}x'_ij_{i_n}
(\rvec',t_0)\rmd^3x'+\left[\nablav^n||\frac{\rmd}{\rmd t}\psf^{(n+1)}(t_0)\right]_i\\
\fl&=&-n\d_{i_1\dots i_n}\intl_{\cal D}x'_{i_1}\dots x'_{i_{n-1}}(x'_ij_{i_n}-
x'_{i_n}j_i)\rmd^3x'-n\d_{i_1\dots i_n}\intl_{\cal D}x'_{i_1}\dots x'_{i_n}\,j_i\rmd^3x'\\
\fl&+&\left[\nablav^n||\frac{\rmd}{\rmd t}\psf^{(n+1)}(t_0)\right]_i,
\eeqan
that is
\beqa\label{22}
\fl a_i^{(n)}=-\frac{n}{n+1}\eps_{kii_n}\d_{i_n}\d_{i_1\dots i_{n-1}}
\intl_{\cal D}x'_{i_1}\dots x'_{i_{n-1}}(\rvec'\times \jvec)_k\rmd^3x'
+\frac{1}{n+1}\left[\nablav^n||\frac{\rmd}{\rmd t}\psf^{(n+1)}(t_0)\right]_i.
\eeqa
By introducing the "vectorial product" $\Tsf^{(n)}\times\avec$ as the $n$th order tensor with components
$$\left(\Tsf^{(n)}\times\avec\right)_{i_1\dots i_n}=\eps_{i_nij}T_{i_1\dots i_{n-1}i}
a_j$$
and observing that, particularly,
$$\left(\bbox{b}^n\times\avec\right)_{i_1\dots i_n}=b_{i_1}\dots b_{i_{n-1}}
(\bbox{b}\times\avec)_{i_n},$$
we may use in the equation \eref{22} the definition of the $n$th order magnetic
multipolar momentum
given in \cite{castel}
\begin{equation}\label{23}
\msf^{(n)}(t)=\frac{n}{n+1}\intl_{\cal D}\rvec^n\times\jvec(\rvec,t)\rmd^3x.
\end{equation}
So, the equation \eref{22} may be written as
\beqa\label{24}
\fl \avec^{(n)}&=&-\evec_i\eps_{ii_nk}\d_{i_n}\d_{i_1\dots i_{n-1}}\sf{M}_
{i_1\dots i_{n-1},k}(t_0)+\frac{1}{n+1}\nablav^n||\frac{\rmd}{\rmd t}\psf^{(n+1)}(t_0)
\nonumber\\
\fl &=&-\nablav\times\left(\nablav^{n-1}||\msf^{(n)}(t_0)\right)+\frac{1}{n+1}\nablav^n
||\frac{\rmd}{\rmd t}\psf^{(n+1)}(t_0)
\eeqa
Going back to the expansion \eref{18},
\beqan
\widetilde{\Avec}_{rad}(\rvec,t)&=&\frac{\mu_0}{4\pi r}\nablav\times\suml^
\infty_{n=1}\frac{(-1)^{n-1}}{n!}\nablav^{n-1}||\msf^{(n)}(t-\frac{r}{c})\\
&+&\frac{\mu_0}{4\pi r}\suml^\infty_{n=0}\frac{(-1)^n}{(n+1)!}\nablav^n||\frac
{\d}{\d t}\psf^{(n+1)}(t-\frac{r}{c})
\eeqan
and, finally,
\beqa\label{25}
\widetilde{\Avec}_{rad}(\rvec,t)&=&\frac{\mu_0}{4\pi r}\nablav\times\suml^
\infty_{n=1}\frac{(-1)^{n-1}}{n!}\nablav^{n-1}||\msf^{(n)}(t-\frac{r}{c})\nonumber\\
&+&\frac{\mu_0}{4\pi r}\suml^\infty_{n=1}\frac{(-1)^{n-1}}{n!}\nablav^{n-1}||\frac
{\d}{\d t}\psf^{(n)}(t-\frac{r}{c})
\eeqa
Now, we extract from the equation \eref{25} the terms contributing to the radiation.
Because
$$\d_{i_1\dots i_n}f(t-\frac{r}{c})=\frac{(-1)^n}{c^n}\nu_{i_1}\dots \nu_{i_n}
\frac{\rmd^n}{\rmd t^n}f(t-\frac{r}{c})+O(\frac{1}{r}),$$
we have in the first sum of the equation \eref{25}:
\beqan
\fl\nablav\times\left[\nablav^{n-1}||\msf^{(n)}(t-\frac{r}{c})\right]&=&\evec_i
\eps_{ijk}\d_j\d_{i_1\dots i_{n-1}}\sf{M}_{i_1\dots i_{n-1}k}(t-\frac{r}{c})\\
\fl&=&\frac{(-1)^n}{c^n}\evec_i\eps_{ijk}\nu_j\nu_{i_1}\dots \nu_{i_{n-1}}\frac{\rmd^n}
{\rmd t^n}\sf{M}_{i_1\dots i_{n-1}k}(t-\frac{r}{c})+O(\frac{1}{r})\\
\fl&=&\frac{(-1)^{n-1}}{c^n}\left[\nuvec^{n-1}||\frac{\rmd^n}{\rmd t^n}\msf^{(n)}(
t-\frac{r}{c})\right]\times\nuvec+O(\frac{1}{r})\\
\fl&=&\frac{(-1)^{n-1}}{c^n}\nuvec^{n-1}
||\left[\frac{\rmd^n\msf^{(n)}(t-r/c)}{\rmd t^n}\times \nuvec\right]+O)\frac{1}{r})
\eeqan
obtaining, finally, the formula
\beqa\label{26}
\Avec_{rad}(\rvec,t)&=&\frac{\mu_0}{4\pi}\frac{1}{r}\suml^\infty_{n=1}
\frac{1}{n!c^n}\left[\nuvec^{n-1}||\frac{\rmd^n}{\rmd t^n}\msf^{(n)}(t_0)\right]\times
\nuvec\nonumber\\
&+&\frac{\mu_0c}{4\pi}\frac{1}{r}\suml^\infty_{n=1}\frac{1}{n!c^n}\nuvec^{n-1}||
\frac{\rmd^n}{\rmd t^n}\psf^{(n)}(t_0)
\eeqa
giving the explicit contribution of each multipole to the radiation field.
\par It is possible to express $\Avec_{rad}$ by the reduced multipolar tensors (total
symmetric and traceless tensors) by applying the procedure given in \cite{cvsc,cvjpa}. This
will be done in the following section.

\section{Expressing the radiation by reduced multipolar tensors}
For the following calculations it is suitable to write the expansion 
\eref{26} in the more explicit form
\beqa\label{27}
\fl\frac{4\pi r}{\mu_0}\Avec_{rad}&=&\evec_i\suml_{n\geq 1}\frac{1}{n!c^n}
\eps_{ikl}\left[\nuvec^{n-1}||\frac
{\rmd^n}{\rmd t^n}\bbox{\sf{M}}^{(n)}(t_0)\right]_k\nu_l
+\evec_i\suml_{n\geq 1}\frac{1}{n!c^{n-1}}\left[\nuvec^{n-1}||
\frac{\rmd^n}{\rmd t^n}\psf^{(n)}(t_0)\right]_i\nonumber\\
\fl&=& \evec_i\suml_{n\geq 1}\frac{1}{n!c^n}\eps_{ikl}\nu_l\nu_{i_1}
\dots\nu_{i_{n-1}}\frac{\rmd^n}{\rmd t^n}
{\sf M}_{i_1\dots i_{n-1}k}
+\evec_i\suml_{n\geq 1}\frac{1}{n!c^{n-1}}\nu_{i_1}\dots\nu_{i_{n-1}}
\frac{\rmd^n}{\rmd t^n}{\sf P}_{i_1\dots i_{n-1}i}\nonumber\\
\fl
\eeqa
In \cite{cvsc,cvjpa} was done a general procedure for the reduction of 
multipole tensors represented by Cartesian 
  components to symmetric traceless ones in the static and dynamic 
cases. The transformations implied by this reduction are defined such that 
the electromagnetic potentials $\Avec$ and $\Phi$ are modified  
only by gauge transformations implying a specific feature of the dynanmic 
case: the redefinitions of the 
multipole tensors in the lower $k < n$ orders induced by the reduction of 
tensors in a given order $n$.  
In the present paper this procedure is applied to the radiation field and, 
obviously, only the vector potential is to be considered.
\par The reduction of multipole tensors begining with a given order $n$ is 
achieved by the following steps.
\par 1. The reduction of the magnetic $n$th-order tensor $\msf^{(n)}$, 
given by the equation \eref{23}, to a symmetric tensor $\msfsim$. Since 
the magnetic tensor $\msf^{(n)}$ is symmetric only in the first $n-1$ 
indices,  the reduction to a symmetric one may be performed by the 
transformation \cite{cvsc}
\beqa\label{28}
\fl{\sf M}_{i_1\dots i_n}\rightarrow \mcsim
&=&\frac{1}{n}\left[{\sf M}_{i_1\dots i_n}+{\sf M}_{i_ni_2\dots i_{n-1}i_1}+
\dots +{\sf M}_{i_1\dots i_ni_{n-1}}\right]\nonumber\\
\fl&=&{\sf M}_{i_1\dots i_n}-\frac{1}{n}\suml^{n-1}_{\la=1}
\left[{\sf M}_{i_1\dots i_n}
-{\sf M}_{i_1\dots i_{\la-1}i_{\la+1}\dots i_{n-1}i_ni_{\la}}\right]\nonumber\\
\fl&=& {\sf M}_{i_1\dots i_n}-\frac{1}{n}\suml^{n-1}_{\la=1}\left[{\sf M}_{i_1\dots i_{\la-1}i_{\la+1}\dots i_{n-1}i_{\la}i_n}-
{\sf M}_{i_1\dots i_{\la-1}i_{\la+1}\dots i_{n-1}i_ni_{\la}}\right]\nonumber\\
\fl&=&{\sf M}_{i_1\dots i_n}-\frac{1}{n}\suml^{n-1}_{\la=1}\left[
{\sf M}^{(\la)}_{i_1\dots i_{n-1}i_{\la}i_n}-{\sf M}^{(\la)}_{i_1\dots 
i_{n-1}i_ni_{\la}}\right]\nonumber\\
\fl&=&{\sf M}_{i_1\dots i_n}-\frac{1}{n}\suml^{n-1}_{\la=1}\eps_{i_{\la}i_n q}{\sf N}^
{(\la)}
_{i_1\dots i_{n-1}q}
\eeqa
where we use the notations
\begin{equation}\label{29}
{\sf N}_{i_1\dots i_{n-1}}=\eps_{i_{n-1}ps}{\sf M}_{i_1\dots i_{n-2}ps},
\;\;\;
f^{(\la)}_{i_1\dots i_n}=f_{i_1\dots i_{\la-1}i_{\la+1}\dots i_n}.
\end{equation}
If $\msf^{(n)}$ is given by the original definition \eref{23}, the 
$n-1$th-order tensor $\Nsf^{(n-1)}$ is given by
\beqa\label{30}
{\sf N}_{i_1\dots i_{n-1}}&=&
\frac{n}{n+1}\intl_{\cal D}\xi_{i_1}\dots \xi_{i_{n-2}}\eps_{i_{n-1}ps}
\xi_p(\xivec\times 
\jvec)_s\rmd^3\xi\nonumber\\
&=&\frac{n}{n+1}\intl_{\cal D}\xi_{i_1}\dots \xi_{i_{n-2}}\left[\xivec\times 
(\xivec\times\jvec)\right]_{i_{n-1}}\rmd^3\xi.
\eeqa
We write explicitly the modification of the potential $\Avec_{\rad}$ induced 
by the substitution \eref{28}:
\beqan
\fl  \frac{4\pi r}{\mu_0}\Avec_{\rad}&\rightarrow&\frac{4\pi r}{\mu_0}
\Avec_{\rad}-\frac{\evec_i}
{n!nc^n}\eps_{ikl}\nu_l\nu_{i_1}\dots \nu_{i_{n-1}}\suml^{n-1}_{\la=1}
\eps_{i_{\la}kq}\frac{\rmd^n}{\rmd t^n}\sf{N}^{(\la)}_{i_1\dots i_{n-1}q}
(t_0)\\
\fl  &=&\frac{4\pi r}{\mu_0}\Avec_{\rad}-\frac{\nuvec}{n!nc^n}\suml^{n-1}
_{\la=1}\nu^{(\la)}_{i_1\dots i_{n-1}}\nu_q\frac{\rmd^n}{\rmd t^n}
{\sf N}^{(\la)}_{i_1\dots i_{n-1}q}\\
\fl &+&\frac{\evec_i}{n!nc^n}\suml^{n-1}_{\la=1}\nu^{(\la)}_{i_1\dots i_{n-1}}
\frac{\rmd^n}{\rmd t^n}{\sf N}^{(\la)}_{i_1\dots i_{n-1}i}\\
\fl &=& \frac{4\pi r}{\mu_0}\Avec_{\rad}+\frac{n-1}{n!nc^n}\left[\nuvec^{
n-2}||\frac{\rmd^n}{\rmd t^n}\Nsf^{(n-1)}(t_0)\right]
-\frac{n-1}{n!nc^n}\left[\nuvec^{n-1}||\frac{\rmd^n}{\rmd t^n}\Nsf^{(n-1)}
(t_0)\right]\nuvec
\eeqan
or
\begin{equation}\label{31}
\frac{4\pi r}{\mu_0}\Avec_{\rad}\rightarrow \frac{4\pi r}{\mu_0}\Avec_{\rad}
+\frac{n-1}{n!nc^n}\left[\nuvec^{n-2}||\frac{\rmd^n}{\rmd t^n}\Nsf^{(n-1)}(t_0)
\right]+\frac{4\pi r}{\mu_0}\psi(\rvec,t)\nuvec
\end{equation}
where
\begin{equation}\label{32}
\psi(\rvec,t)=-\frac{\mu_0}{4\pi r}\frac{n-1}{n!nc^n}\left[\nuvec^{n-1}||
\frac{\rmd^n}{\rmd t^n}\Nsf^{(n-1)}(t_0)\right]
\end{equation}
and do not contribute to the fields $\Evec_{\rad}$ and $\Bvec_{\rad}$ 
corresponding to a gauge transformation of the potential.
\par 2. The {\it extra-gauge} alteration of the vector potential by the transformation 
\eref{31}
 may be set off by the transformation of the electric multipolar tensor  
$\psf^{(n-1)}$:
\begin{equation}\label{33}
\psf^{(n-1)}\rightarrow \psf^{'(n-1)}=\psf^{(n-1)}-\frac{n-1}{c^2n^2}
\frac{\rmd}{\rmd t}\Nsf^{(n-1)}
\end{equation}
such that the final transformation of the potential is the gauge transformation
\begin{equation}\label{34}
\Avec_{\rad}\rightarrow \Avec_{\rad} +\psi\nuvec.
\end{equation}

\par 3. After the reduction of the magnetic tensor $\msf^{(n)}$ to a 
symmetric one, we have to perform the reduction to a symmetric traceless 
tensor $\widetilde{\msf}^{(n)}$. This reduction is achieved by the 
transformation \cite{cvsc}
\begin{equation}\label{35}
\mcsim\rightarrow \widetilde{\sf M}_{i_1\dots i_n}=\mcsim-\suml_{D(i)}
\delta_{i_1i_2}\Lambda_{i_3\dots i_n}
\end{equation}
where $\latens^{(n-2)}$ is a symmetric tensor and the sum over $D(i)$ is 
the sum over all permutations of the symbols $i_1,\dots, i_n$ which give 
distinct terms.  Applequist \cite{apple} has given  an explicit formula 
for expressing the components of the symmetric traceless tensor $\widetilde{\msf}^{(n)}$ 
in terms of the traces of the tensor $\msf^{(n)}$, ({\it the detracer theorem}, 
\cite{apple}, equation (5.1)):
\beqa\label{36}
\fl \widetilde{{\sf M}}_{i_1\dots i_n}&=&\mcsim\nonumber\\
\fl &-&\suml^{[n/2]}_{m=1}\frac{(-1)^{m-1}
(2n-1-2m)!!}{(2n-1)!!}\suml_{D(i)}\delta_{i_1i_2}\dots \delta_{i_{2m-1}i_{2m}}
{\sf M}^{(n:m)}_{{\scr (sym)}i_{2m+1}\dots i_n}
\eeqa
where $[n/2]$ denotes the integer part of $n/2$ and ${\sf M}^{(n:m)}_{{\scr (sym)}i_{2m+1}
\dots i_n}$  the components of the $(n-2m)$th-order tensor obtained from 
$\msfsim$ by the contractions of $m$ pairs of symbols $i$, and obviously
\beqa\label{37}
 \fl \suml_{D(i)}\delta_{i_1i_2}\dots \delta_{i_{2m-1}i_{2m}}
{\sf M}^{(n:m)}_{{\scr (sym)}i_{2m+1}\dots i_n}=
\frac{1}{2^m(n-2m)!m!}\suml_{P(i)}\delta_{i_1i_2}\dots \delta_{i_{2m-1}i_{2m}}
{\sf M}^{(n:m)}_{{\scr (sym)}i_{2m+1}\dots i_n}\nonumber
\fl\\
\eeqa
where the sum over $P(i)$ is the sum over {\it all} the permutations of 
the symbols $i$. Using equation \eref{36} we may give explicitly the 
components of the tensor $\latens^{(n-2)}$:
\begin{equation}\label{38}
\fl \Lambda_{i_3\dots i_n}=\suml^{[n/2]}_{m=1}\frac{(-1)^{m-1}(2n-1-2m)!!}
{(2n-1)!!\,m}\suml_{D(i)}\delta_{i_3i_4}\dots \delta_{i_{2m-1}i_{2m}}
M^{(n:m)}_{{\scr (sym)}i_{2m+1}\dots i_n}.
\end{equation}
In terms of the tensor $\latens$ the modification of $\Avec_{\rad}$ induced 
by the substitutions \eref{28},\eref{33} and \eref{35} is obtained by a straightforward calculation:
\beqan
\fl \frac{4\pi r}{\mu_0}\Avec_{\rad}&\rightarrow& \frac{4\pi r}{\mu_0}
\Avec_{\rad}+\frac{4\pi r}{\mu_0}\psi\nuvec
-\frac{\evec_i}{n!c^n}\eps_{ikl}\nu_l\nu_{i_1}\dots\nu_{i_{n-1}}
\suml_{D(i)}
\delta_{i_1i_2}\frac{\rmd^n}{\rmd t^n}\Lambda_{i_3\dots i_{n-1}k}\\
\fl &=&\frac{4\pi r}{\mu_0}\Avec_{\rad}+\frac{4\pi r}{\mu_0}\psi\nuvec-
\evec_i\frac{n-1}{n!c^n}
\eps_{ikl}\nu_l\nu_k\nu_{i_1}\dots\nu_{i_{n-2}}\frac{\rmd^n}{\rmd t^n}
\Lambda_{i_1\dots i_{n-2}}\\
\fl &-&\evec_i\frac{(n-1)(n-2)}{2n!c^n}\eps_{ikl}\nu_l\nu_{i_1}\dots 
\nu_{i_{n-3}}\frac{\rmd^n}{\rmd t^n}\Lambda_{i_1\dots i_{n-3}k}\\
\fl &=& \frac{4\pi r}{\mu_0}\Avec_{\rad}+\frac{4\pi r}{\mu_0}\psi\nuvec+
\evec_i\frac{(n-1)(n-2)}{2n!c^n}\eps_{ilk}\nu_l\left[\nuvec^{n-3}||
\frac{\rmd^n}{\rmd t^n}\latens^{(n-2)}\right]_k.
\eeqan
So, the transformation of the potential  may be written as
\begin{equation}\label{39}
\fl \frac{4\pi r}{\mu_0}\Avec_{\rad}\rightarrow \frac{4\pi r}{\mu_0}\Avec_{\rad}
+\frac{4\pi r}{\mu_0}\psi\nuvec+\frac{(n-1)(n-2)}{2n!c^n}\nuvec\times\left[
\nuvec^{n-3}||\frac{\rmd^n}{\rmd t^n}\latens^{(n-2)}\right]
\end{equation}
\par 4. It is a simple matter to see that the last {\it extra-gauge} term  
  may be set off by the transformation
\begin{equation}\label{40}
\msf^{(n-2)}\rightarrow\msf^{'(n-2)}=\msf^{(n-2)}+\frac{n-2}{2c^2n}
\frac{\rmd^2}{\rmd t^2}\latens^{(n-2)}.
\end{equation}
\par 5. This step consists in the reduction of the symmetric $n$th-order 
electric multipolar tensor $\psf^{(n)}$ to a symmetric and traceless one 
by a transformation of the type \eref{35}:
\begin{equation}\label{41}
{\sf P}_{i_1\dots i_n}\rightarrow \widetilde{{\sf P}}_{i_1\dots i_n}=
{\sf P}_{i_1\dots i_n}-\suml_{D(i)}\delta_{i_1i_2}\Pi_{i_3\dots i_n}
\end{equation}
where the symmetric tensor $\pitens^{(n-2)}$ is defined in terms of the traces 
of the tensor $\psf^{(n)}$ by a relation similar to equation \eref{38}. 
The resulting transformation of $\Avec_{\rad}$ is 
\begin{equation}\label{42}
\fl \frac{4\pi r}{\mu_0}\Avec_{\rad}\rightarrow \frac{4\pi r}{\mu_0}
\Avec_{\rad}
+\frac{4\pi r}{\mu_0}\psi\nuvec-\frac{\evec_i}{n!c^{n-1}}\nu_{i_1}\dots
\nu_{i_{n-1}}\frac{\rmd^n}{\rmd t^n}\suml_{D(i)}\delta_{i_1i_2}\Pi
_{i_3\dots i_{n-1}i}.
\end{equation}
In this equation we have $(n-1)$ terms with $\delta_{ik},\;k=1,\dots,n-1$ 
and $(n-1)(n-2)/2$ terms with $\delta_{i_ji_k},\; j,k=1,2,\dots,n-1$  so that 
\beqan
 \frac{4\pi r}{\mu_0}\Avec_{\rad}&\rightarrow& \frac{4\pi r}{\mu_0}
\Avec_{\rad}+\frac{4\pi r}{\mu_0}\psi\nuvec
-\frac{n-1}{n!c^{n-1}}\nuvec\,\nu_{i_2}\dots\nu_{i_{n-1}}\frac{\rmd^n}
{\rmd t^n}\Pi_{i_2\dots i_{n-1}}\\
&-&\frac{(n-1)(n-2)}{2n!c^{n-1}}\nu_{i_3}\dots\nu_{i_{n-1}}
\frac{\rmd^n}{\rmd t^n}\Pi_{i_3\dots i_{n-1}i}
\eeqan
that is
\begin{equation}\label{43}
\frac{4\pi r}{\mu_0}\Avec_{\rad}\rightarrow \frac{4\pi r}{\mu_0}
\Avec_{\rad}+\frac{4\pi r}{\mu_0}(\psi+\psi')\nuvec-\frac{(n-1)(n-2)}
{2n!c^{n-1}}\left[\nu^{n-3}||\frac{\rmd^n}{\rmd t^n}\pitens^{(n-2)}\right]
\end{equation}
where
\begin{equation}\label{44}
\psi'=-\frac{4\pi r}{\mu_0}\frac{(n-1)}{n!c^{n-1}}\nuvec^{n-2}||
\frac{\rmd^n}{\rmd t^n}\pitens^{(n-2)}.
\end{equation}
\par 6. The alteration of the potential 
represented by the last term in equation \eref{43} is set off by the 
transformation 
\begin{equation}\label{45}
\psf^{(n-2)}\rightarrow \psf^{(n-2)}+\frac{n-2}{2nc^2}\frac{\rmd^2}
{\rmd t^2}\pitens^{(n-2)}.
\end{equation}    
which preserves the symmetry proprties of $\psf^{(n-2)}$.
\par By this last transformation \eref{45}, the reduction of the multipolar 
tensors in the given $n$th-order is achieved. Now, to carry out this 
procedure to the $n-1$th order, we must realize that in this order some 
tensors was been already modified in order to set off the alterations of 
the electromagnetic field by the reductions in the $n$th-order. So, the 
transformation  \eref{33} alters the symmetry properties of the $(n-1)$th 
order electric multipole tensor because 
\begin{equation}\label{46}
\delta \psf^{(n-1)}=-\frac{n-1}{c^2n^2}\frac{\rmd}{\rmd t}\Nsf^{(n-1)}(t_0)
\end{equation}
is symmetric only in the first $n-2$ indices. To restore the full symmetry 
of the $(n-1)$th-order electric moment, we perform the reduction of $\Nsf^{(n-1)}$ 
to a symmetric tensor by the transformation
\begin{equation}\label{47}
{\sf N}_{i_1\dots i_{n-1}}\rightarrow {\sf N}_{i_1\dots i_{n-1}}\rightarrow 
{\sf N}_{i_1\dots i_{n-1}}-\frac{1}{n-1}\suml^{(n-2)}_{\la=1}\left[{\sf N}_{i_1\dots i_{n-1}}
{\sf N}^{(\la)}_{i_1\dots i_{n-1}i_{\la}}\right]
\end{equation}
By introducing the tensor ${\cal N}^{(n-2)}$ with the components
\begin{equation}\label{48}
{\cal N}_{i_1\dots i_{n-2}}=\eps_{i_{n-2}ps}{\sf N}_{i_1\dots i_{n-3}ps}
\end{equation}
the transformation \eref{47}may be written as

$${\sf N}_{i_1\dots i_{n-1}}\rightarrow {\sf N}_{i_1\dots i_{n-1}}-\frac{1}{n-1}\suml^{(n-2)}
_{\la=1}\eps_{i_{\la}i_{n-1}q}{\cal N}^{(\la)}_{i_1\dots i_{n-2}q}$$
and
\beqa\label{49}
\fl{\sf P}'_{i_1\dots i_{n-1}}&=&{\sf P}_{i_1\dots i_{n-1}}-\frac{n-1}{c^2n^2}\frac{\rmd}{\rmd t}
{\sf N}_{i_1\dots i_{n-1}}\nonumber\\
&\rightarrow &
{\sf P}_{i_1\dots i_{n-1}}-\frac{n-1}{c^2n^2}\frac{\rmd}{\rmd t}{\sf N}_{i_1\dots i_{n-1}}
+\frac{1}{c^2n^2}\suml^{(n-2)}_{\la=1}
\eps_{i_{\la}i_{n-1}q}\frac{\rmd}{\rmd t}{\cal N}^{(\la)}_{i_1\dots i_{n-2}q}.
\eeqa
If $\msf^{(n)}$ is given by the original definition \eref{23}, then we can 
 write
\begin{equation}\label{50}
{\cal N}_{i_1\dots i_{n-2}}=-\frac{n}{n+1}\intl_{{\cal D}}\xi^2\xi_{i_1}\dots 
\xi_{i_{n-3}}\left(\xivec\times\jvec\right)_{i_{n-2}}\rmd^3\xi.
\end{equation}
The alteration of the vector potential $\Avec$ by the transformation 
\eref{49} is given by
\begin{equation}\label{51}
\Avec\rightarrow \Avec-\frac{n-2}{n!nc^n}\nuvec\times\left[\nuvec^{n-3}\times
\frac{\rmd}{\rmd t}\bbox{{\cal N}}^{(n-2)}(t_0)\right].
\end{equation}
This alteration of $\Avec$ is set  off by the transformation of $\msf^{'(n-2)}$,
 given by equation \eref{40}, 
\begin{equation}\label{52}
\msf^{'(n-2)}\rightarrow\msf^{''(n-2)}=\msf^{'(n-2)}-\frac{n-2}{n^2(n-1)c^2}
\frac{\rmd^2}{\rmd t^2}\bbox{{\cal N}}^{(n-2)}.
\end{equation}
By this transformation the symmetry properties of $\msf^{'(n-2)}$ are 
preserved. Particularly, by reducing the $(n-2)$th-order multipolar tensors, 
in the case of $\msf^{''(n-2)}$ we have to achieve only the symmetrisation 
of the supplementary term from the equation \eref{40}.

\section{Concluding remarks}
The equation \eref{1} is a basic formula in the investigation of the radiation 
of electric charges distributions. In a course on Electrodynamics the 
radiation chapter is one of the mains goals  of the introduction of 
Maxwell's equations. Sometimes it is necessary to give quickly some 
results illustrating the properties of the electromagnetic radiation. 
For the economy of this course it is a benefit to avoid some unnecessary 
intermediate results such as  the expressions of the fields 
 $\Evec$ and $\Bvec$. By exposing this problem in the sections 2 and 3, the 
present paper is not claimed as an original one but would be warranted 
at least by concerns of methodological completeness.
\par We summarize the results of the sections 4 and 5 by the following statements.
\par (1) The reduction of the magnetic $n$th-order multipole tensor to 
a symmetric traceless one by the transformation (equations \eref{28} and 
\eref{35})
\begin{equation}\label{53}
{\sf M}_{i_1\dots i_n}\rightarrow \widetilde{{\sf M}}_{i_1\dots i_n}=
{\sf M}_{i_1\dots i_n}-\frac{1}{n}\suml^{n-1}_{\la=1}\eps_{i_{\la}i_nq}
{\sf N}^{(\la)}_{i_1\dots i_{n-1}q}-\suml_{D(i)}\delta_{i_1i_2}\Lambda_{
i_3\dots i_n}
\end{equation}
together the modifications of the electric $(n-1)$th-order and of the magnetic 
$(n-2)$th-order multipole tensors,
\begin{equation}\label{54}
{\sf P}_{i_1\dots i_n}\rightarrow {\sf P}_{i_1\dots i_n}-\frac{n-1}{c^2n^2}
\dot{{\sf N}}_{i_1\dots i_{n-1}},
\end{equation}
where we use the super dot notation for the time derivatives and,
\begin{equation}\label{55}
{\sf M}_{i_1\dots i_{n-2}}\rightarrow {\sf M}_{i_1\dots i_{n-2}}+
\frac{n-2}{2c^2n}\ddot{\Lambda}_{i_1\dots i_{n-2}},
\end{equation}
leads to a gauge transformation of the potential $\Avec_{\rad}$.
\par We point out that if $\msf^{(n)}$ is given by the equation \eref{23},
 the  symmetric traceless tensor 
$\widetilde{\msf}^{(n)}$ may be identified with  the 
 tensor $\cm^{(n)}$ given by  \cite{cvsc}
\begin{equation}\label{56}
{\cal M}_{i_1\dots i_n}(t)=\frac{(-1)^n}{(n+1)(2n-1)!!}
\suml^n_{\la=1}\intl_{{\cal D}}r^{2n+1}\left[\jvec(\rvec,t)\times\nablav)\right]
_{i_{\la}}\d^{(\la)}_{i_1\dots i_n}\frac{1}{r}\rmd^3x.
\end{equation}
\par (2) The reduction of the electric $n$th-order multipole tensor 
 to a symmetric traceless one ,
\begin{equation}\label{57}
{\sf P}_{i_1\dots i_n}\rightarrow \widetilde{{\sf P}}=
{\sf P}_{i_1\dots i_n}-\suml_{D(i)}\delta{i_1i_2}\Pi_{i_3\dots i_n},
\end{equation}
together with the transformation of the electric $(n-2)$th-order tensor,
\begin{equation}\label{58}
{\sf P}_{i_1\dots i_{n-2}}\rightarrow {\sf P}_{i_1\dots i_{n-2}}+
\frac{n-2}{2nc^2}\ddot{\Pi}_{i_1\dots i_{n-2}},
\end{equation}
leads also to a gauge transformation of the vector potential $\Avec_{\rad}$. 
\par We point out also that if $\psf^{(n)}$ is given by the equation \eref{21}, 
the  symmetric traceless tensor 
$\widetilde{\psf}^{(n)}$ may be identified with the tensor $\cp^{(n)}$ given by 
\cite{jansen}
\begin{equation}\label{59}
{\cal P}_{i_1\dots i_n}=\frac{(-1)^n}{(2n-1)!!}\intl_{{\cal D}}
\rho(\rvec,t)r^{2n+1}\nablav^n\frac{1}{r}\rmd^3x.
\end{equation}
If we begin the reduction from a given order n, then the results of the 
reductions of $\psf^{(n)}$ and $\psf^{(k)}$ are the tensors 
$\bbox{{\cal P}}^{(n)}$  and $\bbox{{\cal M}}^{(n)}$ given by the equations 
\eref{59} and \eref{56} but for $k<n$ the $k$th-order reduced multipole 
tensors may differ from $\bbox{{\cal P}}^{(k)}$ and $\bbox{\cal M}^{(k)}$ 
by terms iduced by the procedure of the reductions from the previous steps. 
These last terms give contributions to the potentials and fields expressed 
by toroidal moments and mean radii of various orders.
\par We give here some simple examples of such reductions and we will see 
how naturally the toroidal moments appear as a result of such an approach. 
\par Let us the reduction of the magnetic and electric multipole tensors 
begins from the $\mu th$  and $\eps th$ orders respectively (generally, 
considering the multipole's contributions of the same orders, $\mu=\eps-1$). 
\par For $(\mu,\eps)=(1,2)$, we have $\msf^{(1)}\rightarrow\widetilde{\msf}
^{(1)}=\msf^{(1)},\;\psf^{(1)}\rightarrow \widetilde{\psf}^{(1)}=
\psf^{(1)},\;\psf^{(2)}\rightarrow\widetilde{\psf}^{(2)}={\mathcal P}^{(2)}$. 
These transformations produce only a gauge transformation of $\Avec$.
\par For $(\mu,\eps)=(2,3),\;(3,4),\;(4,5),\;(5,6)$ the reductions are given in 
Appendix. For arbitrary $\mu$ and $\eps$, we think this is possible  to find a 
general rule or, at least,
to elaborate symbolic computer programs.
\par In the case $(\mu,\eps)=(2,3)$,
\beqa \label{60}
\fl\msf^{(2)}&\rightarrow&\widetilde{
\msf}^{(2)}={\mathcal M}^{(2)},\;\msf^{(1)}\rightarrow\widetilde{\msf}^{(1)}=
\msf^{(1)},\,\psf^{(3)}\rightarrow\widetilde{\psf}^{(3)}={\mathcal P}^{(3)},\;
\psf^{(2)}\rightarrow \widetilde{\psf}^{(2)}={\mathcal P}^{(2)}\nonumber\\
\fl\psf^{(1)}&\rightarrow&\widetilde{\psf}^{(1)}=\psf^{(1)}-\frac{1}{4c^2}\dot{
\Nsf}^{(1)}+\frac{1}{6c^2}\ddot{\Pi}^{(1)}.
\eeqa
Here, $\Nsf^{(1)}$ and $\Pi^{(1)}$ are given by the equation \eref{a10} 
for $\psf_{qqi}=0,\;\Nsf_{qqi}=0,\;\psf_{qqppi}=0$ that is eliminating the 
contributions from the orders $n_\mu > 2$ of the magnetic multipole tensors 
and from the orders $n_{\eps} > 3$ for the electric ones.
\par Taking into account the continuity equation verified by $\rho$ and 
$\jvec$, we obtain 
\begin{equation}\label{61}
\widetilde{\sf P}_i={\sf P}_i-\frac{1}{c^2}\dot{\sf T}_i,
\end{equation}
where 
\begin{equation}\label{62}
{\sf T}_i=\frac{1}{10}\intl_{\mathcal D}\left[(\xivec\cdot\jvec)\xi_i-2\xi^2j_i\right]\rmd^3\xi,
\end{equation}
is the toroid dipole tensor [10-12].
\par In the case $(\mu,\eps)=(3,4)$ we have the changes
\beqa\label{63}
\fl\msf^{(3)}&\rightarrow&\widetilde{\msf}^{(3)}={\mathcal M}^{(3)},\;
\msf^{(2)}\rightarrow\widetilde{\msf}^{(2)}={\mathcal M}^{(2)},\;
\msf^{(1)}\rightarrow\widetilde{\msf}^{(1)}=\msf^{(1)}+\frac{1}{c^2}\ddot
{\Lambda}^{(1)}\nonumber\\
\fl\psf^{(4)}&\rightarrow&\widetilde{\psf}^{(4)}={\mathcal P}^{(4)},\;
\psf^{(3)}\rightarrow\widetilde{\psf}^{(3)}={\mathcal P}^{(3)},\;
\psf^{(2)}\rightarrow\widetilde{\psf}^{(2)}={\mathcal P}^{(2)}-\frac{2}
{9c^2}\dot{\widetilde{\Nsf}}^{(2)}+\frac{1}{4c^2}\ddot{\Pi}^{(2)}\nonumber\\
\fl \psf^{(1)}&\rightarrow&\widetilde{\psf}^{(1)}=\psf^{(1)}-\frac{1}{4c^2}
\dot{\Nsf}^{(1)}+\frac{1}{6c^2}\ddot{\Pi}^{(1)}
\eeqa
where 
\begin{equation}\label{64}
\widetilde{\Nsf}_{ij}=\frac{1}{2}(\Nsf_{ij}+\Nsf_{ji})
\end{equation}
and $\Lambda_i,\;\Nsf_{ij}$ and $\Pi_{ij}$ are given by the equations 
\eref{a3}, \eref{a4}, \eref{a7} and \eref{a8} by eliminating the contributions 
from the orders $n_\mu > 3$ and $n_{\eps} > 4$. In this case one obtains 
the contribution of the toroidal quadrupol tensor $\Tsf^{(2)}$, [10-13]:
\begin{equation}\label{65}
{\sf T}_{ik}=\frac{1}{42}\intl_{\cal D}\left[4(\xivec\cdot\jvec)\xi_i\xi_k-5
\xi^2(\xi_ij_k+\xi_kj_i)+2\xi^2(\xivec\cdot\jvec)\delta_{ik}\right]\rmd^3\xi
\end{equation}
having, beside the equation \eref{61}, 
\begin{equation}\label{66}
\widetilde{\sf P}_{ik}=\mathcal{P}_{ik}-\frac{1}{c^2}\dot{\sf T}_{ik}
\end{equation}
and the  dipolar  magnetic moment modified by a mean-square
current radius:
$$\widetilde{\sf M}_i={\sf M}_i+\frac{1}{c^2}\frac{1}{20}\intl_{\mathcal D}
\xi^2\left(\xivec\times\jvec\right)\rmd^3\xi.$$
\par In the case $(\mu,\eps)=(4,5)$ we obtain the following results of the 
reductions:
\beqa\label{67}
\fl \widetilde{\msf}^{(4)}&=&{\mathcal  M}^{(4)},\;\widetilde{\msf}^{(3)}=
{\mathcal M}^{(3)},\nonumber\\
\fl\widetilde{\msf}^{(2)}&=&{\mathcal M}^{(2)}+\frac{1}{4c^2}\ddot{\Lambda}^{(2)}
-\frac{1}{24c^2}\ddot{\mathcal N}^{(2)}:\nonumber\\
\fl \widetilde{\sf M}_{ik}&=&{\mathcal M}_{ik}+
\frac{1}{42c^2}\intl_{\mathcal D}\xi^2
\left[\xi_i\left(\xivec\times\jvec\right)_k+\xi_k\left(\xivec\times
\jvec\right)_i\right]\rmd^3\xi,\nonumber\\
\fl\widetilde{\sf M}_i&=&{\sf M}_i+\frac{1}{20c^2}\intl_
{\mathcal D}\xi^2\left(\xivec\times\jvec\right)_i\rmd^3\xi;\\
\fl\widetilde{\psf}^{(5)}&=&{\mathcal P}^{(5)},\;\widetilde{\psf}^{(4)}=
{\mathcal P}^{(4)},\nonumber\\
\fl\widetilde{\psf}^{(3)}&=&{\mathcal P}^{(3)}-\frac{3}{16c^2}
\dot{\widetilde{\Nsf}}^{(3)}+\frac{3}{10c^2}\ddot{\widetilde{\Pi}}^{(3)}:
\widetilde{\sf P}_{ijk}={\mathcal P}_{ijk}-\frac{1}{c^2}\dot{\sf T}_{ijk},\nonumber\\
\fl{\sf T}_{ijk}&=&\frac{1}{60}\intl_{\mathcal D}\Big[\xi^4\suml_
{D(i,j,k)}\delta_{ij}j_k
+\xi^2\left(\xivec\cdot\jvec\right)\suml_{D(i,j,k)}\delta_{ij}\xi_k+
5\left(\xivec\cdot\jvec\right)\xi_i\xi_j\xi_k-5\xi^2\suml_{D(i,j,k)}\xi_i\xi_jj_k
\Big]\rmd^3\xi,\nonumber\\
\fl\widetilde{\psf}^{(2)}&=&{\mathcal P}^{(2)}-\frac{2}{9c^2}
\dot{\widetilde{\Nsf}}^{(2)}+\frac{1}{4c^2}\ddot{\widetilde{\Pi}}^{(2)}:\;
{\sf P}_{ij}={\mathcal P}_{ij}-\frac{1}{c^2}\dot{\sf T}_{ij},\nonumber\\
\fl \widetilde{\psf}^{(1)}&=&\psf^{(1)}-\frac{1}{4c^2}\dot{\Nsf}^{(1)}+\frac{1}
{96c^4}\dot{\Nsf}^{'(1)}:\;\widetilde{\sf P}_i={\mathcal P}_i-\frac{1}{c^2}
\dot{\sf T}_i-\frac{1}{c^4}\tdot{\Delta}_i,\nonumber\\
\fl \Delta_i&=&\frac{1}{1400}\intl_{\mathcal D}\big[10\xi^2\left(\xivec\cdot
\jvec\right)\xi_i-15\xi^4j_i\big]\rmd^3\xi\nonumber
\eeqa
\par These results show that one may obtain from the formula \eref{26} 
the correct representation of the electromagnetic field by the reduced 
multipolar tensors but introducing these tensors up to a given order $n$, 
we obtain separate contributions from some electric toroidal moments and 
mean $2n-$power radii. This was pointed out firstly by Dubovik {\it et al} [
10-12]. In the present paper we point out that considering the contributions 
to the electromagnetic field of some toroidal moments, one suppose the 
reduction of the multipole tensors up to a well defined maximal order $n$. 
\\ We illustrate also this statement by calculating the total power radiated 
by a system of electric charges and currents.
\par Let us the total radiation power obtained by  integrating the equation 
\eref{1}:
\begin{equation}\label{68}
  {\cal I}_{\mu,\eps}=\frac{1}{\mu_0c}\int\left(\nuvec\times\dot{\Avec}\right)^2
_{\mu, e}r^2\rmd\Omega(\nuvec)
\end{equation}
considering only the contributions of the magnetic and electric multipoles 
up to the $\mu$th and $\eps$th orders respectively. Usig equation \eref{26} 
we may write
\beqa\label{69}
\fl &~&\left(\frac{4\pi r}{\mu_0}\right)^2\left(\nuvec\times\Avec\right)^
2_{\mu,e}\nonumber\\
\fl &=& \suml^{\mu}_{n=1}\suml^{\mu}_{m=1}\frac{1}{n!m!c^{n+m}}\left[
\left(\nuvec^{n-1}||\msf^{(n)}_{,n}\right)\cdot\left(\nuvec^{m-1}||\msf^{(m)}_{,m}
\right)-\left(\nuvec^n||\msf^{(n)}_{,n}\right)\left(\nuvec^m||\msf^{(m)}_{,m}
\right)\right]\nonumber\\
\fl &+&\suml^{\eps}_{n=1}\suml^{\eps}_{m=1}\frac{1}{n!m!c^{n+m-2}}\left[\left(
\nuvec^{n-1}||\psf^{(n)}_{,n}\right)\cdot\left(\nuvec^{m-1}||\psf^{(m)}_{,m}\right)-
\left(\nuvec^n||\psf^{(n)}_{,n}\right)\left(\nuvec^m||\psf^{(m)}_{,m}\right)\right]
\\
\fl &+& 2\suml^{\mu}_{n=1}\suml^{\eps}_{m=1}\frac{1}{n!m!c^{n+m-1}}\left\{\left(\nuvec^{n-1}||\msf^{(n)}_{,n}\right)\cdot
\left[\nuvec\times\left(\nuvec^{m-1}||\psf^{(m)}_{,m}\right)\right]
\right\}.\nonumber
\eeqa
where $$\Tsf^{(n)}_{,k}=\frac{\rmd^k}{\rmd t^k}\Tsf^{(n)}.$$

The calculation of the integrals in equation \eref{68} is reduced to the 
calculation of $<\nu_{i_1}\dots \nu_{i_n}>_{\nuvec},\;n=0,1,\dots $ 
with 
\beqa \label{70}
\fl<f(\nuvec)>_{\nuvec}&=&\frac{1}{4\pi}\int f(\nuvec)\rmd\Omega(\nuvec),\nonumber\\
\fl&~& <\nu_{i_1}\dots \nu_{i_{2n+1}}>_{\nuvec}=0,\;\;\nonumber\\
\fl&~& <\nu_{i_1}\dots \nu_{i_{2n}}>_{\nuvec}=C_n\suml_{D(i)}
\delta_{i_1i_2}\dots \delta_{i_{2n-1}i_{2n}},\;\;C_n=\frac{1}{(2n+1)!!}
\eeqa
Let us the symmetric traceless tensors  $ \Asf^{(n)}$ and $\Bsf^{(m)}$ and the  
averaged contraction
$$\left<\left(\nuvec^k||\Asf^{(n)}\right)||\left(\nuvec^{k'}||\Bsf^{(m)}\right)
\right>_{\nuvec}=\left<\nu_{i_1}\dots\nu_{i_k}\nu_{j_1}\dots \nu_{j_{k'}}
{\sf A}_{i_1\dots i_ki_{k+1}\dots i_n}B_{j_1\dots {j_{k'}}j_{k'+1}\dots 
j_m}\right>_{\nuvec}. $$
This is non zero only for the products of $\delta_{i_pj_p}$
with  $p=1,\dots k,\;q=1,\dots k'$ and it is easy to demonstrate the relation 

\begin{equation}\label{71}
\left<\left(\nuvec^k||\Asf^{(n)}\right)||\left(\nuvec^{k'}
||\Bsf^{(m)}\right)\right>_{\nuvec}
=\frac{k!}{(2k+1)!!}\left[\Asf^{(n)}||\Bsf^{(m)}\right]\,
\delta_{k'k}.
\end{equation}
The terms of the last sum from the equation \eref{69} give contributions 
to the total radiated power of the form
$$\left<\nu_{i_1}\dots\nu_{i_{n-1}}\nu_{j_1}\dots\nu_{j_{m-1}}\nu_p\right>
\eps_{i_npq}{\sf A}_{i_1\dots i_n}{\sf B}_{j_1\dots j_{m-1}q}$$
but all the terms from the sum of $\delta-$products representing the averaged 
products of $\nu$'s contains either $\delta_{i_kp}$ or $\delta_{pj_l}$, 
$k=1,\dots, n-1,\;l=1,\dots ,m-1$ such that, because of $\eps_{i_npq}$ and 
of the traceless character of ${\Asf}$ and $\Bsf$, the result is zero.
Using these results in equations \eref{68} and \eref{69} we obtain 
\beqa\label{72}
\fl&~&{\mathcal I}_{\mu,\eps}=\\
\fl&~&\frac{1}{4\pi\eps_0c^3}\left[\suml^\mu_{n=1}
\frac{n+1}{nn!(2n+1)!!}\frac{1}{c^{2n}}\left[\widetilde{\msf}^{(n)}_{,n+1}||
\widetilde{\msf}^{(n)}_{,n+1}\right]
 +
\suml^{\eps}_{n=1}
\frac{n+1}{nn!(2n+1)!!}\frac{1}{c^{2n-2}}\left[\widetilde{\psf}^{(n)}_{,n+1}||
\widetilde{\psf}^{(n)}_{,n+1}\right]\right]\nonumber
\eeqa
For comparison with results existing in literature [1,10-12] we write here 
the results in the following cases.
The case $(\mu,\eps)=(1,2)$ is given in \cite{land}:
\beqa\label{73}
{\cal I}_{1,2}&=&\frac{1}{4\pi\eps^3_0}\left[\frac{2}{3}\widetilde
{\psf}^{(1)}_{,2}||\widetilde{\psf}^{(1)}_{,2}+\frac{1}{20c^2}\widetilde
{\psf}^{(2)}_{,3}||\widetilde{\psf}^{(2)}_{,3}+\frac{2}{3c^2}\widetilde
{\msf}^{(1)}_{,2}||\widetilde{\msf}^{(1)}_{,2}\right]\nonumber\\
&=&\frac{1}{4\pi\eps_0c^3}\left[\frac{2}{3}\ddot{\pvec}^2+
\frac{2}{3c^2}\ddot{\mvec}^2+\frac{1}{20c^2}\tdot{\cal P}^{(2)}||
\tdot{\cal P}^{(2)}\right],
\eeqa
this result being justified by the  invariance of the radiation field 
to the transformation $\psf^{(2)}\rightarrow {\cal P}^{(2)}$.
\par In  the case $(\mu,\eps)=(2,3)$ we  obtain
\beqa\label{74}
\fl&&{\cal I}_{2,3}\nonumber\\
\fl&=&\frac{1}{4\pi\eps^3_0}\left[\frac{2}{3}\widetilde
{\psf}^{(1)}_{,2}||\widetilde{\psf}^{(1)}_{,2}+\frac{1}{20c^2}\widetilde
{\psf}^{(2)}_{,3}||\widetilde{\psf}^{(2)}_{,3}+\frac{2}{945c^4}\widetilde
{\psf}^{(3)}_{,4}||\widetilde{\psf}^{(3)}_{,4}+
\frac{2}{3c^2}\widetilde{\msf}^{(1)}_{,2}||\widetilde{\msf}^{(1)}_{,2}
+\frac{1}{20c^4}\widetilde{\msf}^{(2)}_{,3}||\widetilde{\msf}^{(2)}_{,3}
\right]\nonumber\\
\fl&=&\frac{1}{4\pi\eps_0 c^3}\Big[\frac{2}{3}|\ddot
{\pvec}-\frac{1}{c^2}\tdot{\Tsf}|^2+\frac{2}{3c^2}\ddot{\mvec}^2
+\frac{1}{20c^2}\tdot{\mathcal P}^{(2)}||\tdot{\mathcal P}^{(2)}\nonumber\\
\fl&+&\frac{1}{20c^4}\tdot{\mathcal M}^{(2)}||\tdot{\mathcal M}^{(2)}+
\frac{2}{945 c^4}{\mathcal P}^{(3)}_{,4}||{\mathcal P}^{(3)}_{,4}
\Big]
\eeqa
In the case $(\mu,\eps)=(3,4)$,
\beqa\label{75}
\fl&&{\cal I}_{3,4}\nonumber\\
\fl&=&\frac{1}{4\pi\eps^3_0}\left[\frac{2}{3}\widetilde
{\psf}^{(1)}_{,2}||\widetilde{\psf}^{(1)}_{,2}+\frac{1}{20c^2}\widetilde
{\psf}^{(2)}_{,3}||\widetilde{\psf}^{(2)}_{,3}+\frac{2}{945c^4}\widetilde
{\psf}^{(3)}_{,4}||\widetilde{\psf}^{(3)}_{,4}+
\frac{1}{18144c^6}\widetilde{\psf}^{(4)}_{,5}||\widetilde{\psf}^{(4)}_{,5}\right.
\nonumber\\
\fl&+&\left.\frac{2}{3c^2}\widetilde{\msf}^{(1)}_{,2}||\widetilde{\msf}^{(1)}_{,2}
+\frac{1}{20c^4}\widetilde{\msf}^{(2)}_{,3}||\widetilde{\msf}^{(2)}_{,3}
+\frac{2}{945c^4}\widetilde{\msf}^{(3)}_{,4}||\widetilde{\msf}^
{(3)}_{,4}\right]\nonumber\\
\fl&=&\frac{1}{4\pi\eps_0c^3}
\left[\frac{2}{3}\left(\ddot{\pvec}-\frac{1}{c^2}\tdot{\Tsf}\right)^2+
\frac{2}{3c^2}\left(\ddot{\mvec}+\frac{1}{c^2}\frac{\rmd^4}{\rmd t^4}
\Lambda\right)^2\right.\nonumber\\
\fl&+&\frac{1}{20}\left(\tdot{\cal P}^{(2)}-\frac{1}{c^2}\frac{\rmd^4}
{\rmd t^4}{\sf T}^{(2)}\right)||\left(\tdot{\cal P}^{(2)}-\frac{1}{c^2}
\frac{\rmd^4}{\rmd t^4}{\sf T}^{(2)}\right)
+\frac{1}{20}\tdot{\cal M}^{(2)}||\tdot{\cal M}^{(2)}\nonumber\\
\fl &+&\frac{2}{945c^4}\left(\frac{\rmd^4}{\rmd t^4}{\cal P}^{(3)}||
\frac{\rmd^4}{\rmd t^4}{\cal P}^{(3)}
+\frac{1}{c^2}\frac{\rmd^4}{\rmd t^4}{\cal M}^{(3)}||
\frac{\rmd^4}{\rmd t^4}{\cal M}^{(3)}\right)\nonumber\\
\fl&+&\left.\frac{1}{18144\,c^6}\frac{\rmd^5}{\rmd t^5}{\cal P}^{(4)}||
\frac{\rmd^5}{\rmd t^5}{\cal P}^{(4)}\right]
\eeqa

\appendix
\section{Reduction of multipole tensors}
In tis Appendix we give in a diagram form the reductions of multipole tensors 
for the cases $(\mu,\eps)=(2,3),\;(3,4),\;(4,5),\;(5,6)$. 
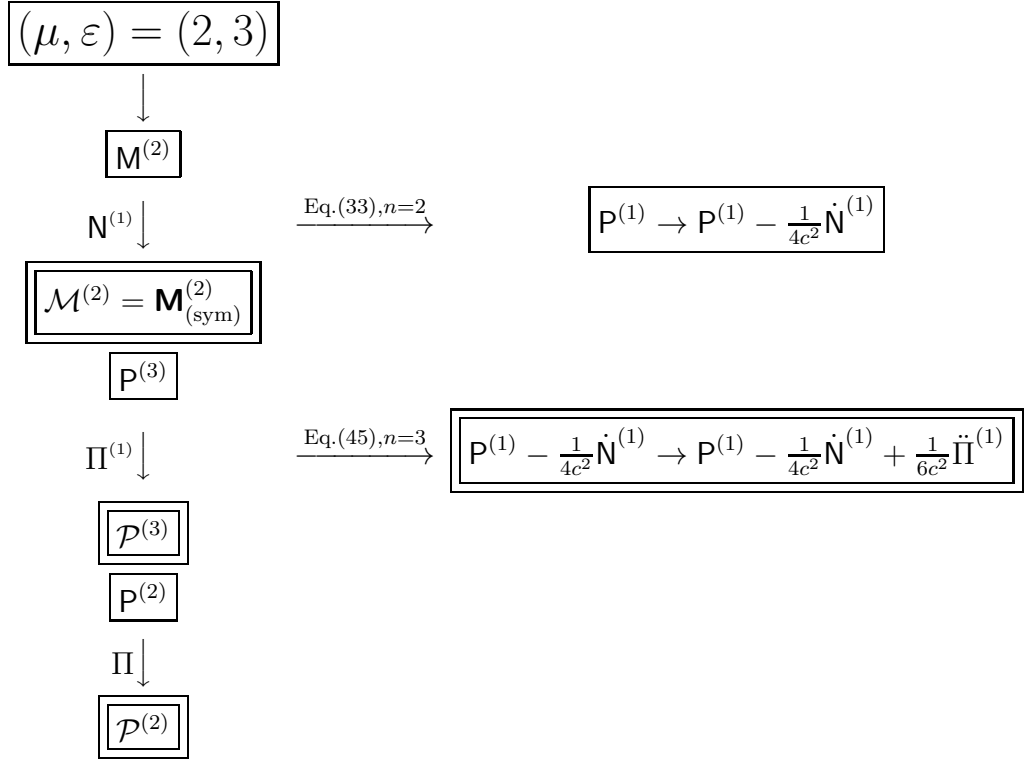
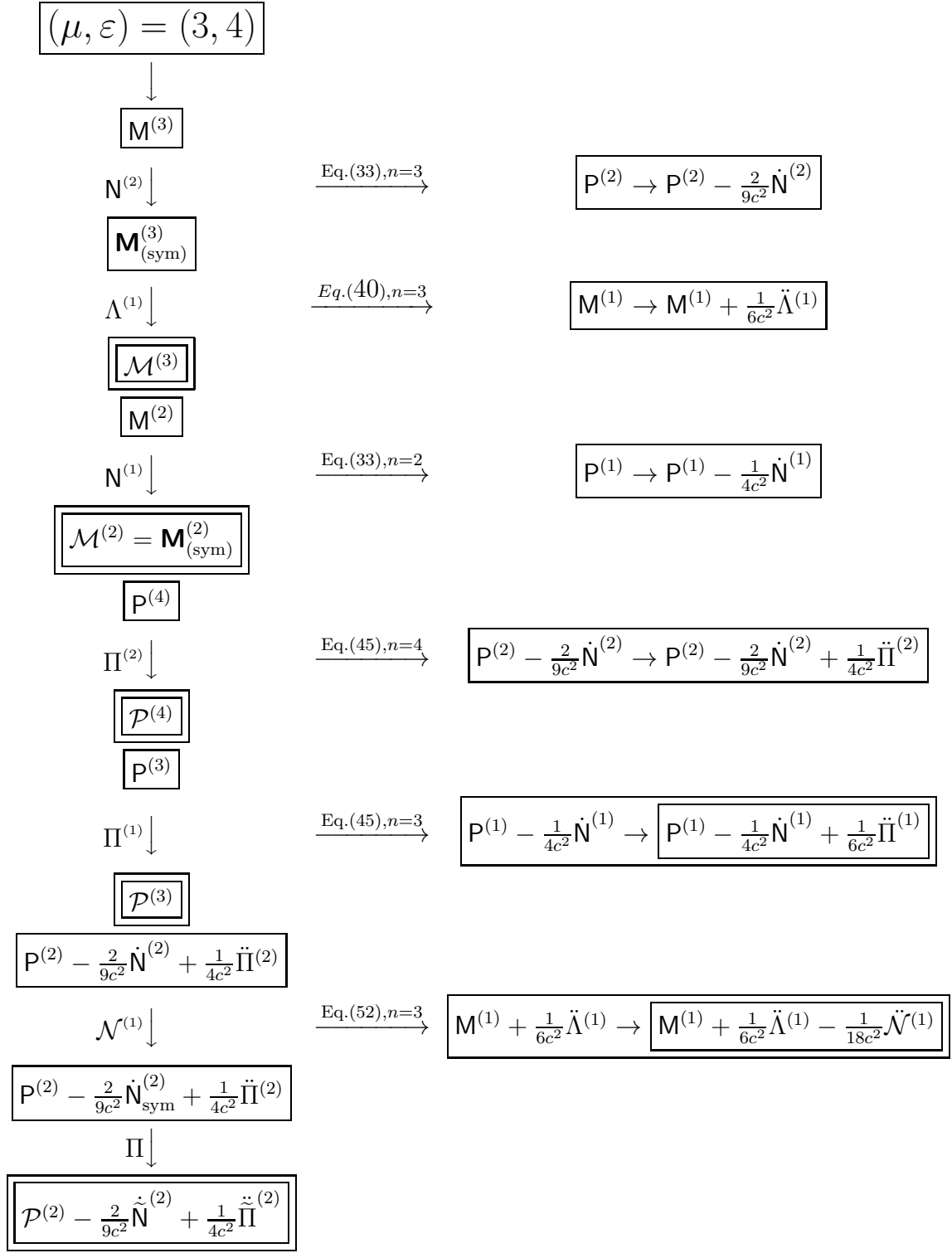
\beqan
\begin{CD}
\fl\framebox{\Large{$(\mu,\eps)=(2,3)$}}\\
\fl@VVV\\
\fl\framebox{$\msf^{(2)}$}\\
\fl @V\Nsf^{(1)}VV @>\mbox{\scriptsize Eq.\eref{33}},
n=2>>
\framebox{$\psf^{(1)}\rightarrow 
\psf^{(1)}-\frac{1}{4 c^2}\dot{\Nsf}^{(1)}$}\\
\fl\framebox{\framebox{${\mathcal M}^{(2)}=\msfsim^{(2)}$}}\\
\fl\framebox{$\psf^{(3)}$}\\
\fl @V\hbox{$\Pi$}^{(1)}VV @>\mbox{\scriptsize Eq.\eref{45}},
n=3>>
\framebox{\framebox{$\psf^{(1)}-\frac{1}{4 c^2}\dot{\Nsf}^{(1)}\rightarrow 
\psf^{(1)}-\frac{1}{4 c^2}\dot{\Nsf}^{(1)}+\frac{1}{6c^2}\ddot
{\hbox{$\Pi$}}^{(1)}$}}\\
\fl\framebox{\framebox{$\mathcal{P}^{(3)}$}}\\
\fl\framebox{$\psf^{(2)}$}\\
\fl @V\hbox{$\Pi$}VV \\
\fl\framebox{\framebox{$\mathcal{P}^{(2)}$}}\\
\end{CD}
\eeqan

\beqan
\begin{CD}
\fl\framebox{\Large{$(\mu,\eps)=(3,4)$}}\\
\fl @VVV\\
\fl\framebox{$\msf^{(3)}$}\\
\fl @V\Nsf^{(2)}VV @>\mbox{\scriptsize Eq.\eref{33}},
n=3>>
\framebox{$\psf^{(2)}\rightarrow 
\psf^{(2)}-\frac{2}{9 c^2}\dot{\Nsf}^{(2)}$}\\
\fl\framebox{$\msfsim^{(3)}$}\\
\fl @V\hbox{$\Lambda$}^{(1)}VV @>Eq.\mbox{\scriptsize}\eref{40},n=3>>
\framebox{$\msf^{(1)}\rightarrow 
\msf^{(1)}+\frac{1}{6 c^2}\ddot{\Lambda}^{(1)}$}\\
\fl\framebox{\framebox{$\mathcal{M}^{(3)}$}}\\
\fl\framebox{$\msf^{(2)}$}\\
\fl @V\Nsf^{(1)}VV @>\mbox{\scriptsize Eq.\eref{33}},
n=2>>
\framebox{$\psf^{(1)}\rightarrow 
\psf^{(1)}-\frac{1}{4 c^2}\dot{\Nsf}^{(1)}$}\\
\fl\framebox{\framebox{${\mathcal M}^{(2)}=\msfsim^{(2)}$}}\\
\fl\framebox{$\psf^{(4)}$}\\
\fl @V\hbox{$\Pi$}^{(2)}VV @>\mbox{\scriptsize Eq.\eref{45}},
n=4>>
\framebox{$\psf^{(2)}-\frac{2}{9 c^2}\dot{\Nsf}^{(2)}\rightarrow 
\psf^{(2)}-\frac{2}{9 c^2}\dot{\Nsf}^{(2)}+\frac{1}{4c^2}\ddot{\hbox
{$\Pi$}}^{(2)}$}\\
\fl\framebox{\framebox{$\mathcal{P}^{(4)}$}}\\
\fl\framebox{$\psf^{(3)}$}\\
\fl @V\hbox{$\Pi$}^{(1)}VV @>\mbox{\scriptsize Eq.\eref{45}},
n=3>>
\framebox{$\psf^{(1)}-\frac{1}{4 c^2}\dot{\Nsf}^{(1)}\rightarrow $
\framebox{$\psf^{(1)}-\frac{1}{4 c^2}\dot{\Nsf}^{(1)}+\frac{1}{6c^2}
\ddot{\hbox{$\Pi$}}^{(1)}$}}\\
\fl\framebox{\framebox{$\mathcal{P}^{(3)}$}}\\
\fl\framebox{$\psf^{(2)}-\frac{2}{9 c^2}\dot{\Nsf}^{(2)}+
\frac{1}{4c^2}\ddot{\Pi}^{(2)}$}\\
\fl @V\hbox{$\mathcal{N}$}^{(1)}VV @>\mbox{\scriptsize Eq.\eref{52}},
n=3>>
\framebox{$\msf^{(1)}+\frac{1}{6c^2}\ddot{\Lambda}^{(1)} \rightarrow $
\framebox{$\msf^{(1)}+\frac{1}{6c^2}\ddot{\Lambda}^{(1)}
-\frac{1}{18c^2}\ddot{\mathcal N}^{(1)}$}}\\
\fl\framebox{$\psf^{(2)}-\frac{2}{9 c^2}\dot{\Nsf}^{(2)}_{\scriptstyle 
\rm{sym}}+\frac{1}{4c^2}\ddot{\Pi}^{(2)}$}\\
\fl @V\hbox{$\Pi$}VV \\
\fl\framebox{\framebox{$\mathcal{P}^{(2)}-\frac{2}{9 c^2}\dot{\widetilde{\Nsf}}^
{(2)}+\frac{1}{4c^2}\ddot{\widetilde{\Pi}}^{(2)}$}}\\
\end{CD}
\eeqan
\beqan
\begin{CD}
\fl\framebox{\Large{$(\mu,\eps)=(4,5)$}}\\
\fl @VVV\\
\fl\framebox{$\msf^{(4)}$}\\
\fl @V\Nsf^{(3)}VV @>\mbox{\scriptsize Eq.\eref{33}},
n=4>>
\framebox{$\psf^{(3)}\rightarrow 
\psf^{(3)}-\frac{3}{16 c^2}\dot{\Nsf}^{(3)}$}\\
\fl\framebox{$\msfsim^{(4)}$}\\
\fl @V\hbox{$\Lambda$}^{(2)}VV @>Eq.\mbox{\scriptsize}\eref{40},n=4>>
\framebox{$\msf^{(2)}\rightarrow 
\msf^{(2)}+\frac{1}{4 c^2}\ddot{\Lambda}^{(2)}$}\\
\fl\framebox{\framebox{$\mathcal{M}^{(4)}$}}\\
\fl\framebox{$\msf^{(3)}$}\\
\fl @V\Nsf^{(2)}VV @>\mbox{\scriptsize Eq.\eref{33}},
n=3>>
\framebox{$\psf^{(2)}\rightarrow 
\psf^{(2)}-\frac{2}{9 c^2}\dot{\Nsf}^{(2)}$}\\
\fl\framebox{$\msfsim^{(3)}$}\\
\fl @V\hbox{$\Lambda$}^{(1)}VV @>Eq.\mbox{\scriptsize}\eref{40},n=3>>
\framebox{$\msf^{(1)}\rightarrow 
\msf^{(1)}+\frac{1}{6 c^2}\ddot{\Lambda}^{(1)}$}\\
\fl\framebox{\framebox{$\mathcal{M}^{(3)}$}}\\
\fl\framebox{$\msf^{(2)}+\frac{1}{4c^2}\ddot{\Lambda}^{(2)}$}\\
\fl @V\Nsf^{(1)}VV @>\mbox{\scriptsize Eq.\eref{33}},
n=2>>
\framebox{$\psf^{(1)}\rightarrow 
\psf^{(1)}-\frac{1}{4 c^2}\dot{\Nsf}^{(1)}$}\\
\fl\framebox{$\msfsim^{(2)}+\frac{1}{4c^2}\ddot{\Lambda}^{(2}$}\\
\fl\framebox{$\psf^{(5)}$}\\
\fl @V\hbox{$\Pi$}^{(3)}VV @>\mbox{\scriptsize Eq.\eref{45}},
n=5>>
\framebox{$\psf^{(3)}-\frac{3}{16c^2}\dot{\Nsf}^{(3)}\rightarrow 
\psf^{(3)}-\frac{3}{16 c^2}\dot{\Nsf}^{(3)}+\frac{3}{10c^2}\ddot{\Pi}^{(3)}$}\\
\fl\framebox{\framebox{$\mathcal{P}^{(5)}$}}\\
\fl\framebox{$\psf^{(4)}$}\\
\fl @V\hbox{$\Pi$}^{(2)}VV @>\mbox{\scriptsize Eq.\eref{45}},
n=4>>
\framebox{$\psf^{(2)}-\frac{2}{9 c^2}\dot{\Nsf}^{(2)}\rightarrow 
\psf^{(2)}-\frac{2}{9 c^2}\dot{\Nsf}^{(2)}+\frac{1}{4c^2}\ddot{\hbox
{$\Pi$}}^{(2)}$}\\
\fl\framebox{\framebox{$\mathcal{P}^{(4)}$}}\\
\fl\framebox{$\psf^{(3)}-\frac{3}{16 c^2}\dot{\Nsf}^{(3)}+
\frac{3}{10c^2}\ddot{\Pi}^{(3)}$}\\
\fl @V\hbox{$\mathcal{N}$}^{(2)}VV @>\mbox{\scriptsize Eq.\eref{52}},
n=4>>
\framebox{$\mathcal{M}^{(2)}+\frac{1}{4c^2}\ddot{\Lambda}^{(2)} \rightarrow 
\mathcal{M}^{(2)}+\frac{1}{4c^2}\ddot{\Lambda}^{(2)}
-\frac{1}{24c^2}\ddot{\mathcal N}^{(2)}$}\\
\fl\framebox{$\psf^{(3)}-\frac{3}{16 c^2}\dot{\Nsf}^{(3)}_{\scriptstyle 
\rm{sym}}+\frac{3}{10c^2}\ddot{\Pi}^{(3)}$}\\
\fl @V\hbox{$\Pi$}^{(1)}VV @>\mbox{\scriptsize Eq.\eref{45}},
n=3>>
\framebox{$\psf^{(1)}-\frac{1}{4 c^2}\dot{\Nsf}^{(1)}\rightarrow 
\psf^{(1)}-\frac{1}{4 c^2}\dot{\Nsf}^{(1)}+\frac{1}{6c^2}\ddot{\hbox
{$\Pi$}}^{(1)}$}\\
\fl\framebox{\framebox{$\mathcal{P}^{(3)}-\frac{3}{16 c^2}\dot{\widetilde{\Nsf}}^
{(3)}+\frac{3}{10c^2}\ddot{\widetilde{\Pi}}^{(3)}$}}\\
\end{CD}
\eeqan
\beqan
\begin{CD}
\fl\framebox{$\psf^{(2)}-\frac{2}{9 c^2}\dot{\Nsf}^{(2)}+
\frac{1}{4c^2}\ddot{\Pi}^{(2)}$}\\
\fl @V\hbox{$\mathcal{N}$}^{(1)}VV @>\mbox{\scriptsize Eq.\eref{52}},
n=3>>
\framebox{$\msf^{(1)}.. \rightarrow $
\framebox{$\msf^{(1)}+\frac{1}{6c^2}\ddot{\Lambda}^{(1)}
-\frac{1}{18c^2}\ddot{\mathcal N}^{(1)}$}}\\
\fl\framebox{$\psf^{(2)}-\frac{2}{9 c^2}\dot{\Nsf}^{(2)}_{\scriptstyle 
\rm{sym}}+\frac{1}{4c^2}\ddot{\Pi}^{(2)}$}\\
\fl @V\hbox{$\Pi$}VV \\
\fl\framebox{\framebox{$\mathcal{P}^{(2)}-\frac{2}{9 c^2}\dot{\widetilde{\Nsf}}^
{(2)}+\frac{1}{4c^2}\ddot{\widetilde{\Pi}}^{(2)}$}}\\
\fl\framebox{$\mathcal{M}^{(2)}+\frac{1}{4c^2}\ddot{\Lambda}^{(2)}
-\frac{1}{24c^2}\ddot{\mathcal N}^{'(2)}$}\\
\fl @V\Nsf^{'(1)}VV @>\mbox{\scriptsize Eq.\eref{33}},n=2>>
\framebox{$\psf^{(1)}..\rightarrow$
\framebox{$\psf^{(1)}-\frac{1}{4 c^2}\dot{\Nsf}^{(1)}+
\frac{1}{6c^2}\ddot{\hbox{$\Pi$}}^{(1)}+\frac{1}{96c^4}\dot{\Nsf}^{'(1)}$}}\\
\fl\framebox{\framebox{$ \mathcal{M}^{(2)}+\frac{1}{4c^2}
\ddot{\widetilde \Lambda}^{(2)}-\frac{1}{24c^2}\ddot{\mathcal N}
^{(2)}_{\scriptstyle \rm{sym}} $}}\\
\end{CD}
\eeqan

\beqan
\begin{CD}
\fl\framebox{\Large{$(\mu,\eps)=(5,6)$}}\\
\fl @VVV\\
\fl\framebox{$\msf^{(5)}$}\\
\fl @V\Nsf^{(4)}VV @>\mbox{\scriptsize Eq.\eref{33}},n=5>>
\framebox{$\psf^{(4)}\rightarrow 
\psf^{(4)}-\frac{4}{25 c^2}\dot{\Nsf}^{(4)}$}\\
\fl\framebox{$\msfsim^{(5)}$}\\
\fl @V\hbox{$\Lambda$}^{(3)}VV @>Eq.\mbox{\scriptsize}\eref{40},n=5>>
\framebox{$\msf^{(3)}\rightarrow 
\msf^{(3)}+\frac{3}{10 c^2}\ddot{\Lambda}^{(3)}$}\\
\fl\framebox{\framebox{$\mathcal{M}^{(5)}$}}\\
\fl\framebox{$\msf^{(4)}$}\\
\fl @V\Nsf^{(3)}VV @>\mbox{\scriptsize Eq.\eref{33}},
n=4>>
\framebox{$\psf^{(3)}\rightarrow 
\psf^{(3)}-\frac{3}{16 c^2}\dot{\Nsf}^{(3)}$}\\
\fl\framebox{$\msfsim^{(4)}$}\\
\fl @V\hbox{$\Lambda$}^{(2)}VV @>Eq.\mbox{\scriptsize}\eref{40},n=4>>
\framebox{$\msf^{(2)}\rightarrow 
\msf^{(2)}+\frac{1}{4 c^2}\ddot{\Lambda}^{(2)}$}\\
\fl\framebox{\framebox{$\mathcal{M}^{(4)}$}}\\
\fl\framebox{$\msf^{(3)}+\frac{3}{10c^2}\ddot{\Lambda}^{(3)}$}\\
\fl @V\Nsf^{(2)}VV @>\mbox{\scriptsize Eq.\eref{33}},
n=3>>
\framebox{$\psf^{(2)}\rightarrow 
\psf^{(2)}-\frac{2}{9 c^2}\dot{\Nsf}^{(2)}$}\\
\fl\framebox{$\msfsim^{(3)}+\frac{3}{10c^2}\ddot{\Lambda}^{(3}$}\\
\fl @V\hbox{$\Lambda$}^{(1)}VV @>Eq.\mbox{\scriptsize}\eref{40},n=3>>
\framebox{$\msf^{(1)}\rightarrow 
\msf^{(1)}+\frac{1}{6 c^2}\ddot{\Lambda}^{(1)}$}\\
\fl\framebox{$\mathcal{M}^{(3)}+\frac{3}{10c^2}\ddot{\widetilde \Lambda}^{(3)}$}\\
\fl\framebox{$\msf^{(2)}+\frac{1}{4c^2}\ddot{\Lambda}^{(2)}$}\\
\fl @V\Nsf^{(1)}VV @>\mbox{\scriptsize Eq.\eref{33}},
n=2>>
\framebox{$\psf^{(1)}\rightarrow 
\psf^{(1)}-\frac{1}{4 c^2}\dot{\Nsf}^{(1)}$}\\
\fl\framebox{$\msfsim^{(2)}+\frac{1}{4c^2}\ddot{\Lambda}^{(2}$}\\
\fl\framebox{$\psf^{(6)}$}\\
\fl @V\hbox{$\Pi$}^{(4)}VV @>\mbox{\scriptsize Eq.\eref{45}},
n=6>>
\framebox{$\psf^{(4)}-\frac{4}{25c^2}\dot{\Nsf}^{(4)}\rightarrow 
\psf^{(4)}-\frac{4}{25 c^2}\dot{\Nsf}^{(4)}+\frac{1}{3c^2}\ddot{\Pi}^{(4)}$}\\
\fl\framebox{\framebox{$\mathcal{P}^{(6)}$}}\\
\fl\framebox{$\psf^{(5)}$}\\
\fl @V\hbox{$\Pi$}^{(3)}VV @>\mbox{\scriptsize Eq.\eref{45}},
n=5>>
\framebox{$\psf^{(3)}-\frac{3}{16c^2}\dot{\Nsf}^{(3)}\rightarrow 
\psf^{(3)}-\frac{3}{16 c^2}\dot{\Nsf}^{(3)}+\frac{3}{10c^2}\ddot{\Pi}^{(3)}$}\\
\fl\framebox{\framebox{$\mathcal{P}^{(5)}$}}\\
\end{CD}
\eeqan
\beqan
\begin{CD}
\fl\framebox{$\psf^{(4)}-\frac{4}{25 c^2}\dot{\Nsf}^{(4)}+
\frac{1}{3c^2}\ddot{\Pi}^{(4)}$}\\
\fl @V\hbox{$\mathcal{N}$}^{(3)}VV @>\mbox{\scriptsize Eq.\eref{52}},
n=5>>
\framebox{$\mathcal{M}^{(3)}..\rightarrow 
\mathcal{M}^{(3)}+\frac{3}{10c^2}\ddot{\widetilde \Lambda}^{(3)}
-\frac{3}{100c^2}\ddot{\mathcal N}^{(3)}$}\\
\fl\framebox{$\psf^{(4)}-\frac{4}{25 c^2}\dot{\Nsf}^{(4)}_{\scriptstyle 
\rm{sym}}
+\frac{1}{3c^2}\ddot{\Pi}^{(4)}$}\\
\fl @V\hbox{$\Pi$}^{(2)}VV @>\mbox{\scriptsize Eq.\eref{45}},
n=4>>
\framebox{$\psf^{(2)}-\frac{2}{9 c^2}\dot{\Nsf}^{(2)}\rightarrow 
\psf^{(2)}-\frac{2}{9 c^2}\dot{\Nsf}^{(2)}+\frac{1}{4c^2}\ddot{\hbox
{$\Pi$}}^{(2)}$}\\
\fl\framebox{\framebox{$\mathcal{P}^{(4)}-\frac{4}{25 c^2}\dot{\widetilde{\Nsf}}^{(4)
}+\frac{1}{3c^2}\ddot{\widetilde{\Pi}}^{(4)}$}}\\
\fl\framebox{$\psf^{(3)}-\frac{3}{16 c^2}\dot{\Nsf}^{(3)}+
\frac{3}{10c^2}\ddot{\Pi}^{(3)}$}\\
\fl @V\hbox{$\mathcal{N}$}^{(2)}VV @>\mbox{\scriptsize Eq.\eref{52}},
n=4>>
\framebox{$\mathcal{M}^{(2)}.. \rightarrow 
\mathcal{M}^{(2)}+\frac{1}{4c^2}\ddot{\Lambda}^{(2)}
-\frac{1}{24c^2}\ddot{\mathcal N}^{(2)}$}\\
\fl\framebox{$\psf^{(3)}-\frac{3}{16 c^2}\dot{\Nsf}^{(3)}_{\scriptstyle 
\rm{sym}}+\frac{3}{10c^2}\ddot{\Pi}^{(3)}$}\\
\fl @V\hbox{$\Pi$}^{(1)}VV @>\mbox{\scriptsize Eq.\eref{45}},
n=3>>
\framebox{$\psf^{(1)}-\frac{1}{4 c^2}\dot{\Nsf}^{(1)}\rightarrow$ 
\framebox{$\psf^{(1)}-\frac{1}{4 c^2}\dot{\Nsf}^{(1)}+\frac{1}{6c^2}\ddot{\hbox
{$\Pi$}}^{(1)}$}}\\
\fl\framebox{\framebox{$\mathcal{P}^{(3)}-\frac{3}{16 c^2}\dot{\widetilde{\Nsf}}^
{(3)}+\frac{3}{10c^2}\ddot{\widetilde{\Pi}}^{(3)}$}}\\
\fl\framebox{$\psf^{(2)}-\frac{2}{9 c^2}\dot{\Nsf}^{(2)}+
\frac{1}{4c^2}\ddot{\Pi}^{(2)}$}\\
\fl @V\hbox{$\mathcal{N}$}^{(1)}VV @>\mbox{\scriptsize Eq.\eref{52}},
n=3>>
\framebox{$\msf^{(1)}.. \rightarrow 
\msf^{(1)}+\frac{1}{6c^2}\ddot{\Lambda}^{(1)}
-\frac{1}{18c^2}\ddot{\mathcal N}^{(1)}$}\\
\fl\framebox{$\psf^{(2)}-\frac{2}{9 c^2}\dot{\Nsf}^{(2)}_{\scriptstyle 
\rm{sym}}+\frac{1}{4c^2}\ddot{\Pi}^{(2)}$}\\
\fl @V\hbox{$\Pi$}VV \\
\fl\framebox{$\mathcal{P}^{(2)}-\frac{2}{9 c^2}\dot{\widetilde{\Nsf}}^
{(2)}+\frac{1}{4c^2}\ddot{\widetilde{\Pi}}^{(2)}$}\\
\fl\framebox{$\mathcal{M}^{(3)}+\frac{3}{10c^2}\ddot{\widetilde \Lambda}^{(3)}
-\frac{3}{100c^2}\ddot{\mathcal N}^{(3)} $}\\
\fl @V\Nsf^{'(2)}VV @>\mbox{\scriptsize Eq.\eref{33}}>>
\framebox{$\mathcal{P}^{(2)}..\rightarrow
\mathcal{P}^{(2)}-\frac{2}{9 c^2}\dot{\widetilde{\Nsf}}^
{(2)}+\frac{1}{4c^2}\ddot{\widetilde{\Pi}}^{(2)}+\frac{1}{150c^4}
\dot{\Nsf}^{'(2)}$}\\
\fl\framebox{$ \mathcal{M}^{(3)}+\frac{3}{10c^2}\ddot{\widetilde \Lambda}^{(3)}
-\frac{3}{100c^2}\ddot{\mathcal N}^{(3)}_{\scriptstyle \rm{sym}} $}\\
\fl @V\hbox{$\Lambda$}^{'(1)}VV @>Eq.\mbox{\scriptsize}\eref{40}>>
\framebox{$\msf^{(1)}..\rightarrow$
\framebox{$\msf^{(1)}+\frac{1}{6c^2}\ddot{\Lambda}^{(1)}
-\frac{1}{18c^2}\ddot{\mathcal N}^{(1)}-\frac{1}{200c^4}
\ddot{\hbox{$\Lambda$}}^{'(1)}$}}\\
\fl\framebox{\framebox{$\mathcal{M}^{(3)}+\frac{3}{10c^2}\ddot{\widetilde \Lambda}^{(3)}
-\frac{3}{100c^2}\ddot{\widetilde{\mathcal N}}^{(3)} $}}\\
\fl\framebox{$\mathcal{M}^{(2)}+\frac{1}{4c^2}\ddot{\Lambda}^{(2)}
-\frac{1}{24c^2}\ddot{\mathcal N}^{'(1)}$}\\
\fl @V\Nsf^{'(1)}VV @>\mbox{\scriptsize Eq.\eref{33}}>>
\framebox{$\psf^{(1)}..\rightarrow$
\framebox{$\psf^{(1)}-\frac{1}{4 c^2}\dot{\Nsf}^{(1)}+\frac{1}{6c^2}\ddot{\hbox
{$\Pi$}}^{(1)}+\frac{1}{96c^4}\dot{\Nsf}^{'(1)}$}}\\
\fl\framebox{\framebox{$ \mathcal{M}^{(2)}+\frac{1}{4c^2}\ddot{\widetilde \Lambda}^{(2)}
-\frac{1}{24c^2}\ddot{\mathcal N}^{(2)}_{\scriptstyle \rm{sym}} $}}\\
\end{CD}
\eeqan
\beqan
\begin{CD}
\fl\framebox{$\mathcal{P}^{(2)}-\frac{2}{9 c^2}\dot{\widetilde{\Nsf}}^
{(2)}+\frac{1}{4c^2}\ddot{\widetilde{\Pi}}^{(2)}+\frac{1}{150c^2}
\dot{\Nsf}^{'(2)}$}\\
\fl @VVV \\
\fl\framebox{\framebox{$\mathcal{P}^{(2)}-\frac{2}{9 c^2}\dot{\widetilde{\Nsf}}^
{(2)}+\frac{1}{4c^2}\ddot{\widetilde{\Pi}}^{(2)}+\frac{1}{150c^2}
\dot{\Nsf}^{'(2)}_{\scriptstyle \rm{sym}}$}}
\\
\end{CD}
\eeqan
In these diagrams are used the following notations:
\begin{equation}\label{a1}
\fl {\sf N}_{ijkl}=\frac{5}{6}\intl_{\mathcal D}\xi_i\xi_j\xi_k
\left[\xivec\times\left(\xivec\times\jvec\right)\right]_l\rmd^3\xi,
\;{\sf N}_{ijk}=\frac{4}{5}\intl_{\mathcal D}\xi_i\xi_j\left[\xivec\times
\left(\xivec\times\jvec\right)\right]_k\rmd^3\xi
\end{equation}
\begin{equation}\label{a2}
\fl\Lambda_{ijk}=\frac{1}{45}\suml_{D(i)}{\sf M}_{qqijk}
-\frac{1}{14\times 45}\suml_{D(i)}\delta_{ij}{\sf M}_{qqppk}
\end{equation}
\begin{equation}\label{a3}
\fl \Lambda_{ij}=\frac{1}{28}\left({\sf M}_{qqij}+{\sf M}_{qqji}\right),\;\;
{\sf N}_{ik}=\frac{3}{4}\intl_{\mathcal D}\xi_i\left[\xivec\times
\left(\xivec\times \jvec\right)\right]_k\rmd^3\xi,
\end{equation}
\begin{equation}\label{a4}
\fl\Lambda_i=\frac{1}{15}{\sf M}_{qqi}+\frac{3}{5\times 700}
\ddot{\sf M}_{qqppi},\;\;
 {\sf N}_i=\frac{2}{3}\intl_{\mathcal D}\left[\xivec\times\left(\xivec\times
\jvec\right)\right]_i\rmd^3\xi,
\end{equation}
\begin{equation}\label{a5}
\fl\Pi_{ijkl}=\frac{1}{11}{\sf P}_{qqijkl}-\frac{1}{18\times 11}\suml_{D(i)}\delta_{ij}
{\sf P}_{qqppkl}+\frac{1}{21\times 99}\suml_{D(i)}\delta_{ij}\delta_{kl}
{\sf P}_{qqpprr}
\end{equation}
\begin{equation}\label{a6}
\fl \Pi_{ijk}=\frac{1}{9}{\sf P}_{qqijk}-\frac{1}{9\times 14}\suml_{D(i)}
\delta_{ij}{\sf P}_{qqppk},\;\;{\mathcal N}_{ijk}=-\frac{5}{6}\intl_{\mathcal D}
\xi^2\xi_i\xi_j\left(\xivec\times\jvec\right)_k\rmd^3\xi
\end{equation}
\beqa\label{a7}
\fl \Pi_{ij}&=&\left[\frac{1}{7}{\sf P}_{qqij}-\frac{1}{70}\delta_{ij}{\sf 
P}_{qqpp}\right]-\frac{4}{25c^2}\left[\frac{1}{7\times 4}\left(\dot{\sf N}_{qqij}+\dot{\sf N}_{qqji}
\right)\right]\nonumber\\
\fl&+&\frac{1}{3c^2}\left[\frac{1}{7}\ddot{\Pi}_{qqij}-\frac{1}{70}
\delta_{ij}\ddot{\Pi}_{qqpp}\right]
\eeqa
\begin{equation}\label{a8}
\fl {\mathcal N}_{ik}=-\frac{4}{5}\intl_{\mathcal D}\xi^2\xi_i\left(\xivec
\times\jvec\right)_k\rmd^3\xi,\;\;{\mathcal N}_i=-\frac{3}{4}\intl_{\mathcal D}
\xi^2\left
(\xivec\times\jvec\right)_i\rmd^3\xi,\;\;\widetilde{\Pi}_{ij}=\Pi_{ij}
-\frac{1}{3}\delta_{ij}\Pi_{kk}
\end{equation}
\begin{equation}\label{a9}
\fl {\sf N}'_{ik}=-\frac{5}{6}\intl_{\mathcal D}\xi^2\xi_i\left[\xivec\times
\left(\xivec\times\ddot{\jvec}\right)\right]_k\rmd^3\xi,\;\;\Lambda'_k=
\frac{1}{15}\ddot{\mathcal N}_{qqk}=-\frac{1}{18}\intl_{\mathcal D}\xi^4
\left(\xivec\times\ddot{\jvec}\right)_k\rmd^3\xi
\end{equation}
\begin{equation}\label{a10}
\fl {\sf N}'_k=-\frac{4}{5}\intl_{\mathcal D}\xi^2\left[\xivec\times\left(
\xivec\times\ddot{\jvec}\right)\right]_k\rmd^3\xi,\;\;
\Pi_i=\frac{1}{5}{\sf P}_{qqi}-\frac{1}{80c^2}\dot{\sf N}_{qqi}+
\frac{3}{700c^2}\ddot{\sf P}_{qqppi}
\end{equation}
and,
\beqan
\widetilde{\Lambda}_{ijk}&=&\Lambda_{ijk}-\frac{1}{5}\suml_{D(i)}\delta_
{ij}\Lambda_{qqk},\\
\widetilde{\sf N}_{ijkl}&=&{\sf N}_{({\scriptstyle \rm{sym}})ijkl}-
\frac{1}{28}\suml_{D(i)}\delta_{ij}\left({\sf N}_{qqkl}+{\sf N}_{qqlk}
\right),\\
\widetilde{\Pi}_{ijkl}&=&
\Pi_{ijkl}-\suml_{D(i)}
\delta_{ij}\left(\frac{1}{7}
\Pi_{qqkl}-\frac{1}{70}\suml_{D(i)}\delta_{kl}\Pi_{qqpp}\right)\\
\widetilde{\sf N}_{ijk}&=&{\sf N}_{({\scriptstyle \rm{sym}})ijk}-\frac{1}{15}
\suml_{D(i)}\delta_{ij}{\sf N}_{qqk},\\
\widetilde{\Pi}_{ijk}&=&\Pi_{ijk}-\frac{1}{5}\suml_{D(i)}\delta_{ij}
\Pi_{qqk},\\
\widetilde{\sf N}_{ij}&=&\frac{1}{2}\left({\sf N}_{ij}+{\sf N}_{ji}\right),\;\;
\widetilde{\Pi}_{ij}=\Pi_{ij}-\frac{1}{3}\delta_{ij}\Pi_{qq},\\
\widetilde{\mathcal N}_{ijk}&=&{\mathcal N}_{({\scriptstyle \rm{sym}})ijk}-
\frac{1}{15}\suml_{D(i)}\delta_{ij}{\mathcal N}_{qqk}.
\eeqan
For $\mu<5$ and $\eps<6$, the quantities $\Nsf,\,\Lambda,...$ are obtained from 
the above  expressions by eliminating the contributions of the magnetic multipolar 
tensors of orders $n_m>\mu$ and of the the electric ones for orders $n_e>\eps$.
\par We point out that the reduction diagrams described in this appendix are valid 
also in the case of an arbitrary electromagnetic field as may be seen from 
\cite{cvjpa}.

\vspace{0.5cm}
\par {\bf References}

\end{document}